\documentclass[twocolumn,american,english,showkeys,amsmath,amssymb,superscriptaddress,a4]{revtex4}
\usepackage[T1]{fontenc}
\usepackage[active]{srcltx}
\usepackage{babel}
\usepackage{array}
\usepackage{textcomp}                

\usepackage{multirow}
\usepackage{amsmath}
\usepackage{amssymb}
\usepackage{graphicx}
\usepackage{subscript}
\usepackage[unicode=true]
 {hyperref}
\usepackage{breakurl}
\usepackage{color}

\makeatletter

\providecommand{\tabularnewline}{\\}

\@ifundefined{textcolor}{}
{%
 \definecolor{BLACK}{gray}{0}
 \definecolor{WHITE}{gray}{1}
 \definecolor{RED}{rgb}{1,0,0}
 \definecolor{GREEN}{rgb}{0,1,0}
 \definecolor{BLUE}{rgb}{0,0,1}
 \definecolor{CYAN}{cmyk}{1,0,0,0}
 \definecolor{MAGENTA}{cmyk}{0,1,0,0}
 \definecolor{YELLOW}{cmyk}{0,0,1,0}
}

\usepackage{multirow}\usepackage{bm}\usepackage{gensymb}\usepackage{threeparttable}

\newcommand {\kB} {k_{\text{B}}}
\newcommand {\dd} {\text{d}}

\makeatother

\begin{document}

\title{Spontaneous Frenkel pair formation in Zirconium Carbide}

\author{Thomas A. Mellan}

\affiliation{Thomas Young Centre for Theory and Simulation of Materials, Department
of Materials, Imperial College London, Exhibition Road, London SW7
2AZ, United Kingdom}
\email{t.mellan@imperial.ac.uk}

\author{Andrew I. Duff}

\affiliation{STFC Hartree Centre, Scitech Daresbury, Warrington WA4 4AD, United
Kingdom}

\author{Michael W. Finnis}

\affiliation{Thomas Young Centre for Theory and Simulation of Materials, Department
of Physics and Department of Materials, Imperial College London, Exhibition
Road, London SW7 2AZ, United Kingdom}

\date{\today}

\begin{abstract}
With density functional theory we have performed molecular dynamics
simulations of ZrC which displayed spontaneous Frenkel pair formation
at a temperature of $3200\,$K,  some $500$\textdegree{} below the melting point.
To understand this behaviour, rarely seen in equilibrium simulations,
we  quenched and examined a set of  lattices containing a Frenkel pair.  Five
metastable structures were found, and their 
formation energies and electronic properties were studied. Their thermal generation was
found to be facilitated by a reduction of between $0.7$ and $1.5$ eV in formation
energy due to thermal expansion of the lattice. With input from a
quasi-harmonic description of the defect free energy of formation, an ideal
solution model was used to estimate lower bounds on their concentration
as a function of temperature and stoichiometry. At 3000 K (0.81 of the melting temperature) their concentration was estimated to be 
1.2\% per mole  in a stoichiometric crystal, and  0.3\% per mole in a crystal with 10\% per mole of constitutional vacancies. Their contribution
to heat capacity, thermal expansion and bulk modulus was estimated.
\end{abstract}

\date{\today}

\maketitle

%\section{Introduction}

Zirconium carbide is a hard,  corrosion-resistant material, with a high melting point
and  metallic conductivity. The attractiveness of ZrC for aerospace
and nuclear applications has been documented.\citep{Kim2010,Razumovskiy2013,Harrison2015,Justin2011}
To support the experimental investigations of ZrC, particularly for safety critical
applications, we seek to understand an predict some of the properties of this material
at high temperature. In this regard,  thermodynamic measurements are particularly challenging at temperatures approaching $3000\,$K or above, and theoretical predictions are few. Notable exceptions filling some of the gaps are the recent ultra-high temperature measurements
by Savvatimsky \emph{et al.}, and \emph{first principles} simulations
by Duff \emph{et al.} \citep{Duff2015,Savvatimskiy2017}. Nevertheless, even the currently most reliable phase diagram  \citep{FernandezGuillermet1995} is necessarily incomplete and based on  empirical modelling, extrapolating from experimental data, which does not explicitly include data for point defects or their interactions.\citep{Davey2017}
%\reply{1) Davey PhD thesis added. In your original comment you mentioned ptb -- what is this? Was this a reference to prb? Imperial add}

Recent publications provide a hint that a ZrC crystal may contain
a significant concentration of intrinsic Frenkel defects at high temperature.
Calculations by Kim \emph{et al.} and Zhang \emph{et al}. at zero
temperature report low enthalpies of interstitial carbon formation,\citep{Kim2010,Zhang2014}
and the analysis of experiments conducted by Savvatimsky \emph{et
al.}\citep{Savvatimskiy2017} near the melting point attribute heat
capacity behaviour at high temperature to possible Frenkel defects.
In this work, we use temperature-dependent density-functional theory (DFT) calculations
to address the possible formation of intrinsic defects, and deduce consequences
thereof for thermal properties.
%\com{I deleted ``and mechanical'' because I'm not sure we can do this. Please reinstate it if I'm wrong.}
%\reply{2) No problem, probably safer not to overstate.}

The plan of the paper is as follows. In Sec. \ref{subsec:Calculations}
we state technical parameters for numerical calculations, followed
by the setting out of basic equations for the thermodynamic analysis
in Sec. \ref{subsec:Thermodynamics}. Section \ref{subsec:Discovery-of-defects}
details the discovery of defects by molecular dynamics, and describes our systematic
search of the defect configuration space. In Sec. \ref{subsec:Defect-structures}
the calculated structures and stability of defects at $0\,$K is reported.
Section \ref{subsec:Electrons-and-phonons} discusses elementary thermal
excitations in the defective crystal. Section \ref{subsec:Defect-thermodynamics}
reports some temperature dependent properties,
including Frenkel defect concentration, thermal expansion, heat capacity
and bulk modulus. Finally the effect of substoichiometry on Frenkel
pair concentration is discussed. The Appendix contains supplementary
details on the predicted ZrC properties and a derivation of the ideal solution model for
the five
%\com{reminder to check}
 Frenkel defects at variable carbon substoichiometry.

\section{Methods\label{sec: Methods}}

\begin{table}
\caption{Plane-wave kinetic energy cutoff and \textbf{k}-point sampling density
for supercell calculations. Supercell size references diagonal expansion
of basis vectors of the conventional unit cell of perfect ZrC. $^{\dagger}$For
two unbound Frenkel pairs it was necessary to increase the lattice
expansion from $2\times2\times2$ to $4\times2\times2$, in order
to obtain exact phonon frequencies at specific \textbf{q}-points.
The \textbf{q}-point\textbf{ }and\textbf{ k}-point meshes were suitably
adapted for these configurations. \label{tab: Sampling-technical-parameters-cutoff-kpoints}}
\begin{tabular}{lccc}
\hline 
Observable & $E^{\text{kin}}$ (eV) & \textbf{k}-point grid & Supercell size\tabularnewline
\hline 
\hline 
$U_{0}(V)$ & 700 & $12\times12\times12$ & $2\times2\times2$\tabularnewline
$U_{\text{FS}}(V_{0})$ & 700 & $3\times3\times3$ & $4\times4\times4$\tabularnewline
$F_{\text{el}}(V,T)$ & 700 & $12\times12\times12$ & $2\times2\times2$\tabularnewline
$F_{\text{qh}}(V)$ & 700 & $6\times6\times6^{\dagger}$ & $2\times2\times2{}^{\dagger}$\tabularnewline
$E_{\text{MD}}(V,T)$ & 500 & $2\times2\times2$ & $2\times2\times2$\tabularnewline
\hline 
\end{tabular}
\end{table}

\subsection{Calculation\label{subsec:Calculations}}

Langevin \emph{NVT} molecular dynamics (MD) simulations were performed for five temperature-volume points up to the melting point on the ZrC thermal expansion curve.\citep{Duff2015}
%\com{is that what you meant?}
%\reply{3) I've changed  this sentence hopefully to make it clearer, but the main point I was trying to make that NVT MD was performed at five temperatures along the Duff ZrC thermal expansion curve is true.}
MD simulations were run for $15\,$ps at each temperature-volume point, with a friction parameter of $0.1$ THz and time-step of $3$ fs. 

%\com{More clarity here: the reader wonders why run more than one ensemble, and how many members were in each ensemble?}
%\reply{4) NPT ensemble mention deleted as I didn't end up following that up with any points in the   main text.}

Periodic plane-wave density function theory (DFT) calculations were
performed using the VASP software,\citep{Kresse1996,Kresse1996a}
with the local density approximation (LDA) exchange-correlation function.\citep{Perdew1981a}
The projector-augmented wave (PAW) method is used,\citep{Kresse1999}
with 4\emph{s}- and 4\emph{p}-Zr electrons included as valence states.
The \textbf{k}-point sampling mesh and  cutoff kinetic energies
are chosen to obtain free energies converged to better than $1\,$meV/atom and $5\,$meV/defect. \textbf{k}-point mesh sampling and plane-wave cutoff values are summarized in Table \ref{tab: Sampling-technical-parameters-cutoff-kpoints}. The DFT crystal energy $U_{0}(V)$ and the phonon free energy $F_{\text{qh}}(V,T)$
were each calculated on a mesh of 11 volumes spanning the range $12.0\,\text{\r{A}}^{3}/\text{atom}$
to $14.7\,\text{\r{A}}^{3}/\text{atom}$. $F_{\text{qh}}(V,T)$ was computed using quasiharmonic
lattice dynamics as implemented in the PHONOPY code\citep{Togo2015a} with the direct method of calculating the dynamical matrix.\citep{Kresse1995,Parlinski1997,Oba2008}
%\reply{5) Kresse, Togo, and Parlinski direct method citations added}
At each of the 11 volumes, sets of small displacements were applied.
Each configuration required between 40 and 768 displacements
depending on space group. The dynamical matrix was built from these
 forces.\citep{Kresse1996,Kresse1996a,Perdew1981a,Togo2015a} For defect structures, we assumed that thermal expansion of the crystal  can be modelled by a supercell with homogeneous isotropic principal
axis strains. At each volume, internal coordinates were optimized until
force changes were less than $10^{-6}$ eV/\r{A}. Force calculations used
first-order Methfessel-Paxton electron smearing with an electronic
temperature of 0.1 eV.\citep{Methfessel1989}

The electronic free energy, $F_{\text{el}}(V,T)$, was calculated
using the Mermin finite-temperature formulation of DFT,\citep{Mermin1965}
on a mesh of 10 temperatures and 8 volumes, sampled between $V_{\text{eq}}(T=0\,\text{K})$
and $V_{\text{eq}}(T=3800\,\text{K})$. Electron states are self-consistent
to within $10^{-7}$ eV/atom. The $64$
atom supercell used contains $512$ electrons, for which $384$ bands were
sufficient to span all states with significant partial occupation up to the melting
point. 

\subsection{Analysis\label{subsec:Thermodynamics}}

%In this brief summary of thermodynamic analysis based on free energies,
%please assume $\hbar=k_{\text{B}}=1$ unless noted otherwise.

We calculate the Helmholtz free energy 

\[
F=-\kB T\,\text{ln}\,Z\,,
\]
as a sum of three terms

\begin{equation}
F(V,T)=U_{0}(V)+F_{\text{ph}}(V,T)+F_{\text{el}}(V,T)\,.\label{eq: helmholtzFreeEnergy}
\end{equation}
$U_{0}$ is the total energy of static crystal configurations,
$F_{\text{ph}}$ the vibrational free energy, and $F_{\text{el}}$ the contribution of single-electron
thermal excitations as described in the Mermin functional. 

The calculated free energies are fitted by third-degree polynomials,
\begin{equation}
F(V,T)=\sum_{jk}a_{jk}V^{j}T^{k}\,,\label{eq: helmholtzBasis}
\end{equation}
with $j\ge0$, $k\ge1$ and $j+k\le3$. 

The Helmholtz free energy $F$ is Legendre transformed to the Gibbs
free energy $G$ for calculating thermodynamic quantities at constant external pressure  $p$:

\begin{equation}
G(p,T)=\underset{V}{\text{min}}[F(V,T)+pV]\,.\label{eq: gibbsFreeEnergy}
\end{equation}
The volume minimisation is performed on a mesh of temperatures separated
at $2$ K intervals.

For intrinsic defect $i$ in a stoichiometric $2\times2\times2$ supercell,
$G_{i}$ is used to calculate the defect formation energy:

\begin{equation}
\Delta G_{i}^{2\times2\times2}=G_{i}^{\text{2\ensuremath{\times}2\ensuremath{\times}2}}-G_{\text{perfect}}^{\text{2\ensuremath{\times}2\ensuremath{\times}2}}\,.\label{eq: gibbsEnergyOfFormation}
\end{equation}
The spurious interaction between the defect and its periodic images, both electronic and due to the overlapping strain fields,  is accounted for with a \emph{finite-size}
energy correction,

\begin{equation}
\Delta G_{i}^{\text{bulk}}=\Delta G_{i}^{2\times2\times2}+U_{i,\text{FS}}(V_{0})\,.\label{eq: gibbsEnergyFS}
\end{equation}
The finite-size correction $U_{i,\text{FS}}$, is estimated from calculations at the
zero-temperature equilibrium volume,
\begin{equation}
U_{i,\text{FS}}(V_{0})=\underset{n\to\infty}{\text{lim}}U_{i}^{n\times n\times n}(V_{0})-U_{i}^{2\times2\times2}(V_{0}),\label{eq: finiteSizeCorrection}
\end{equation}
where $U_{i}^{n\times n\times n}(V_{0})$ is the energy to form a
defect in a  $n\times n\times n$ supercell. To estimate
$U_{i,\text{FS}}(V_{0})$ in the limit of non-interacting periodic
images, the defect energy is calculated for supercell sizes ranging
from $2\times2\times2$ (Zr\textsubscript{32}C\textsubscript{32})
to $4\times4\times4$ (Zr\textsubscript{256}C\textsubscript{256}),
and linearly extrapolated to the $1/n=0$ dilute limit. 
%\com{add $1/n$ here?}
%\reply{6) Added. I think as you intended?} 

The equilibrium concentration of intrinsic defects has been calculated using the dilute-limit $\Delta G_{i}^{\text{bulk}}$  with an ideal solution model. The model assumes a partition function of the form
\[
Z^{\text{}}=m_{2}^{N_{2}}m_{3}^{N_{3}}\frac{N!}{\left(N-N_{1}-N_{2}-2N_{3}\right)!\,\left(N_{1}+N_{3}\right)!\,N_{2}!\,N_{3}!}\,,
\]
where $m_i$ are degeneracies and $N_i$ are defect numbers, indexed $i={1,\,2,\,3}$ for vacancies, bound pair Frenkels and unbound pair Frenkels respectively. Z gives
\[
n_{2}=(1-n_{1}-n_{2}-2n_{3})m_{2}\,\text{exp}\left(-\frac{\Delta G_{2}}{T}\right)\,,
\]
and
\[
n_{3}=\frac{\left(1-n_{1}-n_{2}-2n_{3}\right){}^{2}}{\left(n_{1}+n_{3}\right)}\,m_{3}\,\text{exp}\left(-\frac{\Delta G_{3}}{T}\right)\,.
\]
for bound and unbound Frenkel pair concentrations. Model derivation and additional details in Appendix.
%\reply{7) I think it is good to have a couple of equations here about the solution model. Although the details are too obvious to publish, I think I would find these basic equations useful if I reading this paper. I don't mind too much, delete if you like though.}

\clearpage{}

\section{Results\label{sec:Results}}
\subsection{Discovery of defects\label{subsec:Discovery-of-defects}}
\begin{center}
\begin{figure}
\begin{centering}
\includegraphics[scale=0.8]{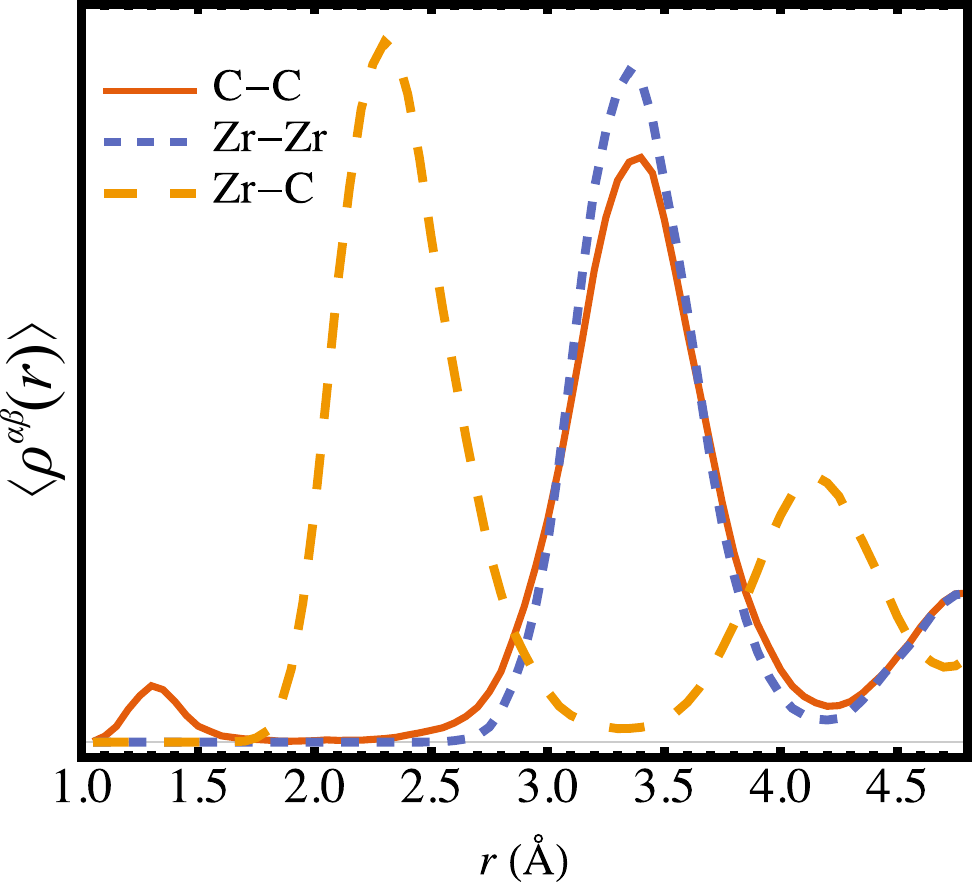}
\par\end{centering}
\caption{Time-averaged pair correlation function, $\left\langle\rho^{\alpha\beta}(r)\right\rangle$ in Eqn. \ref{eq: temporalAveragePairCorrel},
estimated from \emph{ab initio} MD at the melting point $T_{\text{m}}=3700$
K. \label{fig: Pair-correlation-function}}
%\reply{8) DONE- Note, manually edit figure to  include time-averaging brackets in axis label}
\end{figure}
\par\end{center}
\subsubsection*{Molecular dynamics}
With density functional theory (DFT) we first performed \emph{ab initio}
molecular dynamics (AIMD) simulations  for stoichiometric ZrC. Although the timescales accessible in AIMD simulations are
typically insufficient to determine defect statistics such as bound
pair lifetimes, even short runs of 15 ps can provide  insight
into defect behaviour.
Trajectories
were analyzed using an equal-time spatial pair correlation functions defined as follows

%\reply{9) Average number of Zr-Zr, Zr-C, or C-C pairs at separation r at a given instant in time. The number of pairs is counted in a thin shell at r+dr, divided by shell volume 4 pi r2 dr. 
%Thanks for adding the species indices which I had missed. 
%Writing the expression again from the start I think this is okay:}

\begin{equation}
\rho^{\alpha\beta}(r,t)=\frac{1}{N_{\alpha}}\sum_{i,\,j\ne i}\left\langle \delta(r-|r^\alpha_{i}-r^\beta_{j}|)\right\rangle \,,\label{eq: pairCorrelationFunction}
\end{equation}
%\reply{10) I've dropped  the normalisation because I don't think it is necessary but please add if you disagree. And I think the sum should be just over i and j, not r subscript i superscript alpha}
where $\alpha\in \{Zr,C\}$ and $\beta\in \{Zr,C\}$. The time argument is specified in $\rho^{\alpha\beta}(r,t)$ as we do not suppose  the thermodynamic limit for our small ensemble simulations.
%\com{ Two points: 1. Perhaps note that the time dependence of the left hand side is necessary because we are studying the creation of a single Frenkel pair within a small ensemble. In the thermodynamic limit of course the time dependence vanishes.  2. Surely you do need some normalisation because the right hand summation would diverge in an infinite system? }
%\reply{1) Agree, now mentioning the  reason to include (t) in text. 2) My mistake. There is a indeed a factor of $1/N_\alpha$ thanks. Updated in the text. }

Near the melting point, carbon atoms are occasionally observed  to  hop spontaneously
from ($\frac{1}{2}\,\frac{1}{2}\,\frac{1}{2}$) perfect Wyckoff \emph{b}-sites,
with high-symmetry and octahedral coordination, to low symmetry interstitial
sites. Site-resolved pair correlation functions show that newly-formed carbon
interstitials do not immediately leave sites that are nearest-neighbour
to the vacancy they have created.  We therefore refer to the resulting defect as a \emph{bound pair} Frenkel
defect. 

The time-averaged correlation functions,
\begin{align}
\left\langle \rho^{\alpha\beta}(r)\right\rangle &=\frac{1}{\tau}\int\!\!\dd t\,\rho^{\alpha\beta}(r,t)\label{eq: temporalAveragePairCorrel}
\end{align}
are shown for stoichiometric ZrC at the melting point in Fig. \ref{fig: Pair-correlation-function}.
The distinctive carbon-carbon peak at the small separation of $r=1.3$\,\r{A} signifies
interstitial carbon in ZrC. The width of the peak suggests the interstitial carbon is
sufficiently thermally excited to move between multiple coordinations
(which have a range of C-C bond lengths). Examining this behaviour
using AIMD simulations requires significant computational resources. To aid sampling of the interstitial configuration space, interstitial dynamics has been biased to prohibit the recombination of interstitial and vacancy. 
%\com{Is this to stop recombination of interstitial and vacancy? You need to supply more motivation here.}
%\reply{11) Motivation and details added.}

To bias interstitial dynamics a hydrogen atom is placed and frozen
in position at the vacant carbon site. At the melting temperature,
the interstitial is found to exchange between multiple sites within
the vacancy nearest-neighbour (nn) coordination shell. Within $10\,$ps,
the interstitial is observed to begin diffusing, hopping away from
vacancy-nn coordination. Beyond the vicinity of the vacancy-nn, the
displaced carbon can hop between lattice interstitial sites, or switch
places with a perfect-site carbon. 
%In the latter case, the perfect-site
%carbon and original interstitial exchange labels, and the new interstitial
%continues diffusion.
 In AIMD simulations we cannot access fine-grained
diffusion statistics, but note that hopping behaviour between distinct
sites is quite clearly observed on the picosecond timescale.

The Frenkel pairs observed in melting point AIMD are found to be 
metastable at low temperature. This is confirmed by selecting MD configurations
at random, and quenching by steepest-descent to $T=0$ K. The Frenkel
interstitial tends to relax to a C-C-C trimer unit, with a bond angle of $127\text{\textdegree}$  and $C_{2v}$
symmetry. The trimer is found to be stable across
volume dilations ranging from at least -5 \% to +16 \% under homogeneous
principal axis strains.    

\subsubsection*{Systematic search for Frenkel pairs}

In addition to the carbon-trimer of $C_{2v}$ symmetry, identified from quenched
AIMD snapshots, we have systematically searched for other Frenkel
defect configurations with distinct symmetry. 1331
initial Frenkel configurations were considered, each corresponding to a different interstitial position. The interstitial positions were distributed on a uniform grid in a symmetry-reduced wedge of the defective $2\times2\times2$ supercell with other atoms fixed in their perfect lattice sites.
%\com{More precision: Do you mean the interstitial atom was distributed at points on on a uniform grid while other atoms were fixed in their perfect lattice sites?}
%\reply{12) Yes, I did mean that.  I've tried to increase precision.}
Each configuration provided a starting point for a geometry optimization. Optimized configurations were subsequently analyzed in terms of energy and symmetry to identify distinct metastable Frenkel configurations.

In the relaxation of the 1331 initial Frenkel configurations, forces
were finely optimized to better than $10^{-6}$ eV/\r{A}. This was necessary
to prevent relaxations becoming trapped in flat, high-symmetry regions
of the defect configuration space that are not quite local minima.
The high-accuracy relaxations provide data from which we identify
two bound and three unbound interstitial-vacancy pairs that are stable at the equilibrium volume. If $d(C_\text{int}-C_\text{vac})<a/\sqrt{2}$ the vacancy-interstitial pair is classified as bound, and if $d(C_\text{int}-C_\text{vac})>a/\sqrt{2}$ the Frenkel pair is said to be unbound.
%\com{How about: The unbound pairs are defined as those in which the interstitial carbon is not associated with one of the  twelve carbon sites nearest to the vacancy}.
%\reply{13) That is somewhat better but not quite right I think. For example in Fig. 2 d, the interstitial could be associated or bonded with a carbon that is one  of the 12, and it would still be an unbound pair. The definition is, bound pairs have to be in the 12 C nn  box else unbound, not just associated with the box. How about, If $d(C_\text{int}-C_\text{vac})<a/\sqrt{2}$ the vacancy-interstitial pair is defined as bound, and if $d(C_\text{int}-C_\text{vac})>a/\sqrt{2}$ the Frenkel pair is unbound.}

\subsection{Defect structures\label{subsec:Defect-structures}}
\begin{table*}
\caption{This table reports stable Frenkel pair configurations found in ZrC.
Defect configurations are labelled \emph{bound (B)} or \emph{unbound}
\emph{(U)}, with C coordination subscripts. Orientations are specified
by angle brackets which denote the families of directions for the
$C_{\text{i}}-C_{\text{nn}}$ bond in the basis of the direct unitcell
lattice. Orientation indices given are the lowest that provide accuracy
to within 5\% of the exact orientations. The point group refers to the interstitial-carbon
nearest-neighbour unit. Multiplicity $m$ counts the degeneracy of
the interstitial carbon, with respect to vacancies for bound pairs, and per perfect carbon site for unbound pairs. Bond angles and lengths are between interstitial-carbon
and carbon nearest-neighbours.\label{tab: Identified-stable-configurations}}
%\reply{14) Mike reminder - Please double check these multiplicities}
\begin{centering}
\par\end{centering}
\begin{tabular}{ccccccc}
\hline 
Config.  & Description & Orientation & Point group & Multiplicity, $m$\textsubscript{i} & Angle, C\textsubscript{nn}-C\textsubscript{i}-C\textsubscript{nn}
(\textdegree ) & Length, C\textsubscript{nn}-C\textsubscript{i} (\r{A})\tabularnewline
\hline 
\hline 
$B_{2}$ & bound dimer & $\langle332\rangle$ & $D_{\infty h}$ & $24$ & - & 1.411 \tabularnewline
$B_{3}^{\text{}}$ & bound trimer & $\langle654\rangle$ & $C_{2v}$ & $24$ & 126.7 & 1.506 \tabularnewline
$U_{2}$ & unbound dimer & $\langle100\rangle$ & $D_{\infty h}$ & $3$ & - & 1.411 \tabularnewline
$U_{3}^{\text{}}$ & unbound trimer angular & $\langle553\rangle$ & $C_{2v}$ & $12$ & 135.9 & 1.459 \tabularnewline
$U_{3}^{\text{\ensuremath{\text{lin}}}}$ & unbound trimer linear & $\langle110\rangle$ & $C_{\infty v}$ & $6$ & 179.4 & 1.376 \tabularnewline
$U_{4}$ & unbound tetramer  & $\langle322\rangle$ & $C_{3v}$ & $8$ & 117.4 & 1.586 \tabularnewline
\hline 

\end{tabular}

\end{table*}

Of the five defects considered as \emph{bound }(\emph{$B$}) or \emph{unbound}
($U$) Frenkel pairs, the carbon atoms involved are further classified
as dimer, trimer or tetramer, denoted by a subscript suffix. For each
configuration, bound-dimer ($B_{2}$), bound-trimer ($B_{3}$), unbound-dimer
($U_{2}$), unbound-trimer ($U_{3}$), unbound-tetramer ($U_{4}$),
the energetic ordering at the zero-temperature equilibrium volume
is as follows:

\[
perfect<B_{3}<B_{2}<U_{3}<U_{2}<U_{4}\,.
\]

Formation energies are listed in Table \ref{tab:Intrinsic-defect-formation-energy},
each defect structure is pictured in Fig. \ref{fig: structures-defects},
and features of the defect geometries such as bond lengths and point
symmetries are listed in Table \ref{tab: Identified-stable-configurations}.

Note, at the equilibrium volume, the $U_{3}$ configuration is an
angular C-C-C unit, but the linear C-C-C configuration, $U_{3}^{\text{lin}}$,
is stabilised at sufficiently large lattice expansion. Both are shown
in Fig. \ref{fig: structures-defects}.

\begin{center}
\begin{figure*}
\begin{raggedright}
\includegraphics[scale=0.6]{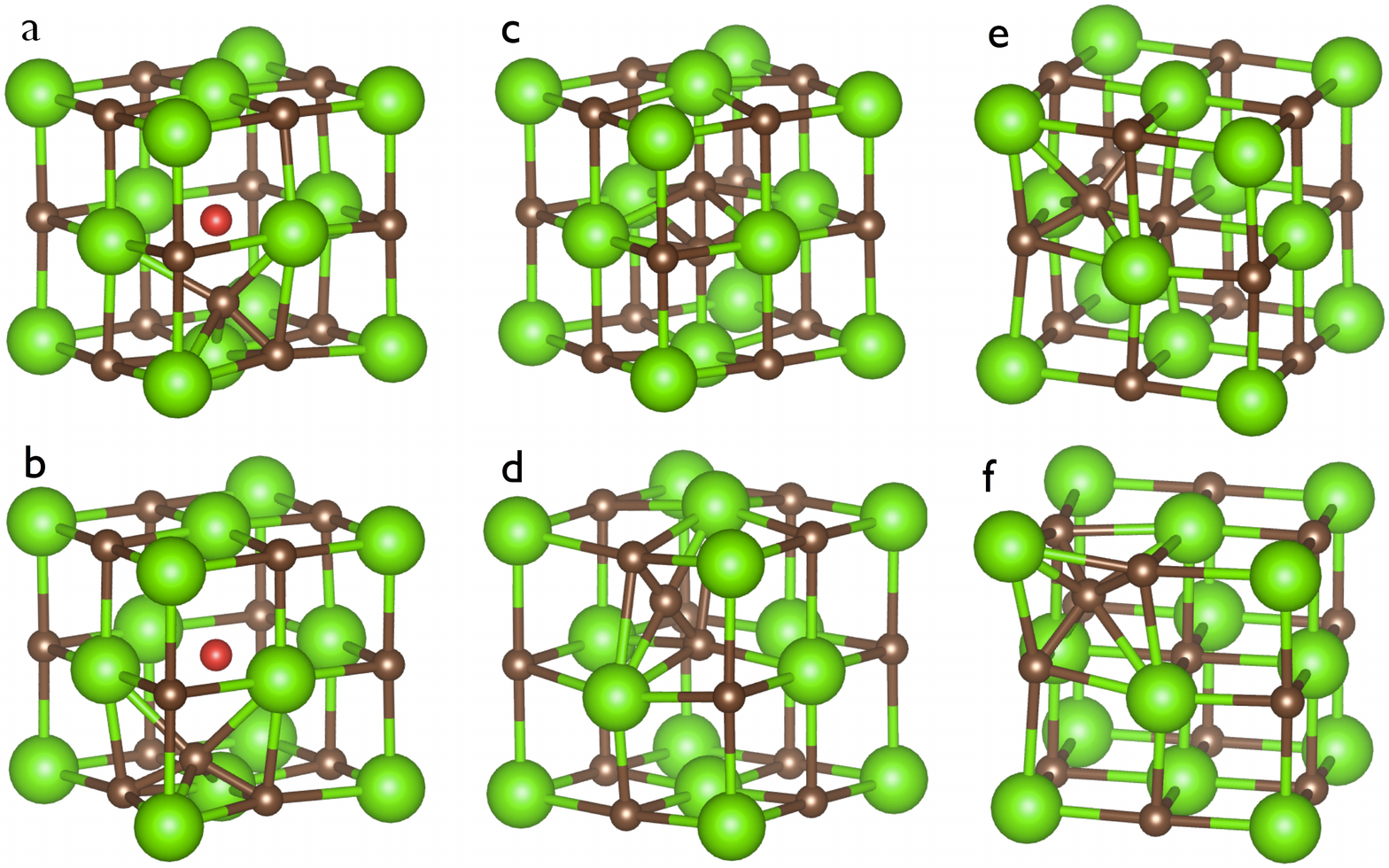}
\par\end{raggedright}
\caption{For each Frenkel pair, the eighth of the supercell that contains the
interstitial atom is visualized.\citep{Momma2011} Carbon atoms are
colored brown, zirconium atoms green, and the vacant carbon site is
highlighted red for bound pair defects. Defect labels are: a) $B_{2}$
bound dimer, b) $B_{3}^{\text{}}$ bound trimer, c) $U_{2}$ unbound
dimer, d) $U_{3}^{\text{lin}}$ unbound linear trimer, e)  $U_{3}$
unbound trimer, and f) $U_{4}$ unbound tetramer.
%\com{Don't you mean  $U_{3}^{\text{lin}}$ should label the unbound linear trimer?}
%\reply{15) Thanks, fixed.}
\label{fig: structures-defects}}
\end{figure*}
\par\end{center}

\subsection{Electrons and phonons\label{subsec:Electrons-and-phonons}}
\begin{center}
\begin{figure*}
\begin{raggedright}
\includegraphics[scale=0.58]{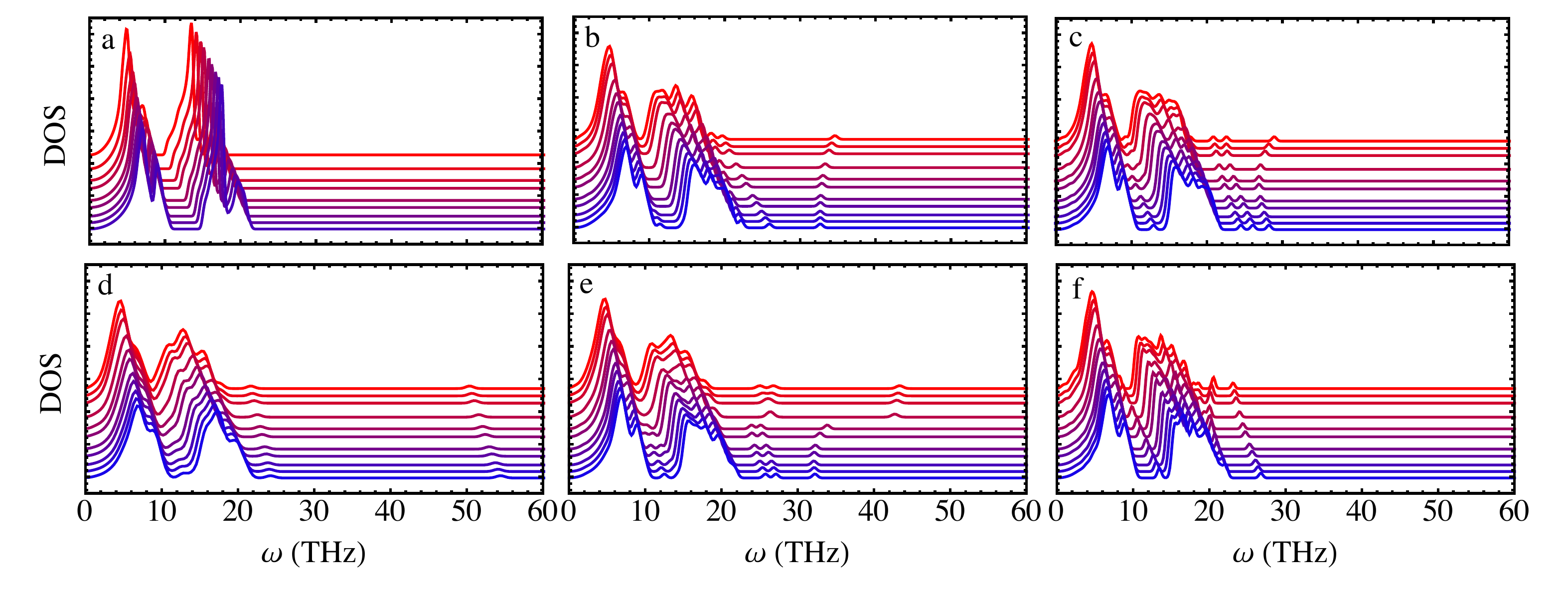}
\par\end{raggedright}
\caption{Phonon density of states for: a) perfect ZrC, b) $B_{2}$ bound dimer,
c) $B_{3}$ bound trimer, d) $U_{2}$ unbound dimer, e) $U_{3}^{\text{}}$
unbound trimer, f) $U_{4}$ unbound tetramer. Thermal expansion increases
from blue to red, along the sequence of volumes \{11.96, 12.16, 12.36,
12.63, 12.85, 13.22, 13.47, 13.83, 14.26, 14.48, 14.70\} \r{A}\protect\textsuperscript{3}/atom.
To highlight Gr{\"u}neisen behaviour, consecutive densities of states
are up-shifted proportional to dilation. \label{fig: qha-phonons} }
\end{figure*}
\par\end{center}

\begin{center}
\begin{figure*}
\begin{raggedright}
\includegraphics[scale=0.56]{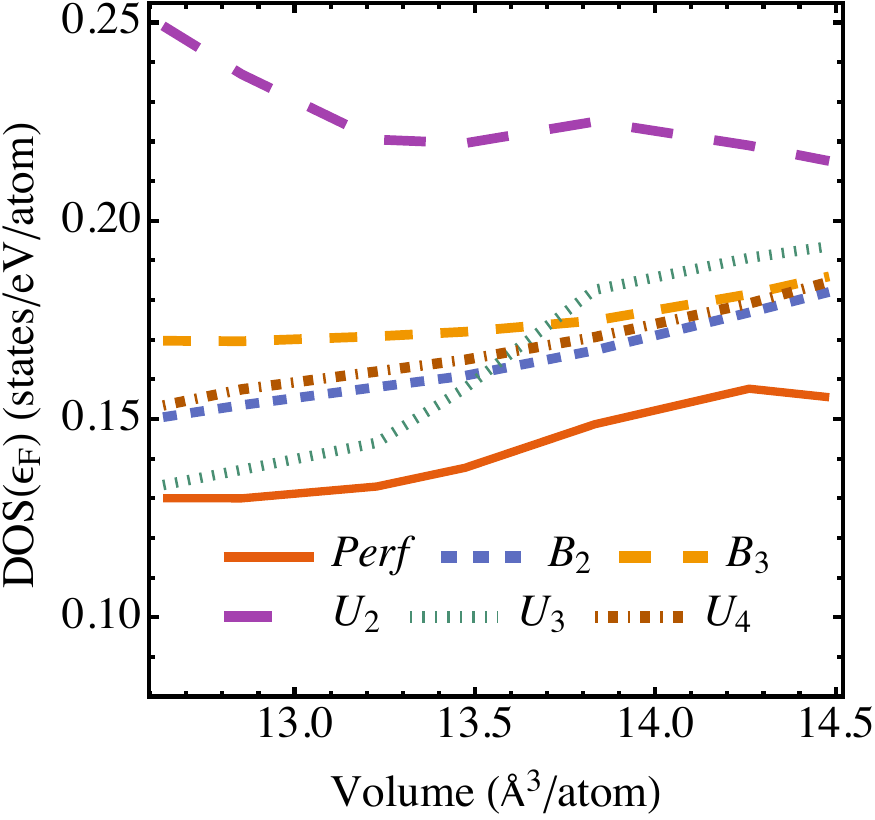}\includegraphics[scale=0.41]{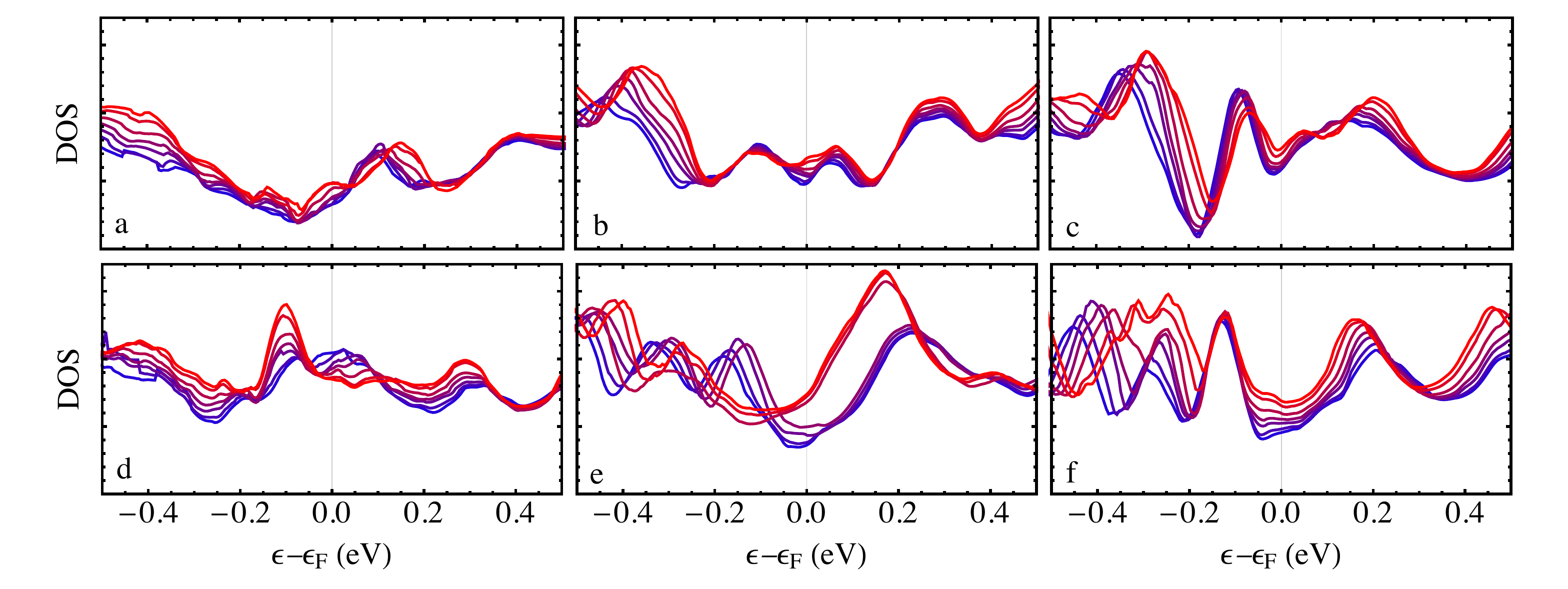}
\par\end{raggedright}
\caption{Electron densities of states (DOS). \emph{LHS}: DOS at the Fermi level
versus crystal expansion. \emph{RHS}: DOS versus energy, with thermal
expansion increasing from blue to red. a) perfect ZrC, b) $B_{2}$
bound dimer, c) $B_{3}$ bound trimer, d) $U_{2}$ unbound dimer,
e) $U_{3}^{\text{}}$ unbound trimer, f) $U_{4}$ unbound tetramer.
Note, electron DOS is provided for on the wider scale of $-6\le\epsilon-\epsilon_{\text{F}}\le6$
in Appendix \ref{sec:Electron-states} Fig. \ref{fig: qha-el-1}.
\label{fig: qha-el} }
\end{figure*}
\par\end{center}

\begin{table}
\caption{Defective to perfect force determinant ratios $\left|\Phi'\right|/\left|\Phi\right|$
and formation entropy for each Frenkel pair. Values are isochoric
at the perfect volume $V_{0}^{\text{perf}}=12.843$ \r{A}/atom. Entropies
resolved to cation and metal contributions. $k_{\text{B}}$/defect
entropy units. \label{tab: defect-Entropy-difference}}

\begin{tabular}{ccccc}
\hline 
Configuration & $\left|\Phi'\right|/\left|\Phi\right|$  & $S^{'}-S$  & $S_{\text{Zr}}^{'}-S_{\text{Zr}}$  & $S_{\text{C}}^{'}-S_{\text{C}}$ \tabularnewline
\hline 
$B_{2}$  & 0.09 & 1.18 & 2.16 & -0.97\tabularnewline
$B_{3}^{\text{}}$  & 0.24 & 0.72 & 2.13 & -1.41\tabularnewline
$U_{2}$  & 0.01 & 2.29 & 2.45 & -0.16\tabularnewline
$U_{3}^{\text{}}$  & 0.06 & 1.40 & 2.53 & -1.13\tabularnewline
$U_{4}$  & 0.08 & 1.26 & 2.46 & -1.20\tabularnewline
\hline 
\end{tabular}
\end{table}

\subsubsection*{Lattice vibrations with defects}

Phonon densities of states (DOS) are shown as a function of frequency
and volume dilation in Fig. \ref{fig: qha-phonons}. Typically vibration
frequencies soften with lattice expansion. For example the mean Gr{\"u}neisen
parameter ($\gamma=-\frac{\partial\text{ln}\bar{\omega}}{\partial\text{ln}V}$)
for perfect ZrC is $\gamma^{\text{perf}}\approx1.5$ at low temperature,
increasing to $\gamma^{\text{perf}}\approx2$ at high temperature.
A notable feature of the perfect ZrC phonon DOS is the obvious gap
in frequencies. For each density of states, below the gap 98.6 \%
of the squared amplitude is on Zr sites, and above the gap 98.6 \%
is on C sites. 

Frenkel defects can be identified in the phonon DOS (Fig. \ref{fig: qha-phonons})
from the distinctive patterns of high frequency vibrations associated
with the interstitial carbon. The patterns shown may provide a useful
spectroscopic fingerprint to experimentally identify Frenkel pairs
in the future.

%\com{You could delete the paragraph about conductivity, and just mention that a study of the effect of these defects on thermal conductivity is the subject of another paper, to be published.}
%\reply{16) Deleted.}

The most stable defect configuration at $T=0$ K is $B_{3}$ (see
Table \ref{tab:Intrinsic-defect-formation-energy}), for which the
phonon DOS is shown in Fig. \ref{fig: qha-phonons}c. The $B_{3}$
defect states comprise localised carbon vibrational frequencies with
one mode red-shifted and three modes blue-shifted enough to lie outside the band of bulk carbon frequencies. The red-shifted defect mode, clearly evident in the gap between
the bulk carbon and zirconium frequencies, is due to rigid C-C-C translation.
For the blue-shifted modes, the highest frequency vibration is an
asymmetric stretch. The other two blue-shifted local modes are types of symmetric stretches, which are distinguished by the motion of the interstitial relative to the local environment as in or out of the C-C-C plane.
%\com{ How can you have \emph{two} symmetric stretches?}
%\reply{17) C-C-C is a planar molecule but the planar symmetry is lost due to the asymmetric environment of the carbon vacancy. I can describe what I observe. Lets label carbon involved in the B3 defect as C1-C2-C3, where C2 is the interstitial. In a symmetric stretch of the angular C-C-C unit, C1 and C3 move along their C1-C2 and C3-C2 bond directions. This has two variations, with C2  moving in the plane of C1-C2-C3, or orthogonal to the C1-C2-C3 plane.  I've tried to change the text to convey this but it is slightly tricky to do so concisely.}

To quantify the high degree of localisation for these
vibrations, consider the highest frequency defect state in Fig. \ref{fig: qha-phonons}c.
For this asymmetric stretch mode, 60.1\% of the motion projects onto
the interstitial, and 19.3\% projects onto each of the two interstitial
nearest neighbours. Only the remaining 1.3\% is not localised on the
immediate C-C-C unit.
For $B_{2}$, shown in Fig. \ref{fig: qha-phonons}b, the blue-shifted
defect frequencies involve a C-C stretching mode which has the highest
frequency, and second mode in which the interstitial rocks about a near-static 
carbon neighbour. The frequency-volume dependence
of the rocking mode mirrors the Gr{\"u}neisen parameter of bulk frequencies,
softening with strain, but the stretching mode anomalously increases
in frequency with lattice expansion.

The $U_{2}$ phonon states in Fig. \ref{fig: qha-phonons}d are further
worth mentioning. The $U_{2}$ DOS shows one remarkably high frequency
peak at $\omega\approx54$ THz. This vibration is due to stretching
of the C-C dimer. The other blue-shifted defect mode is a dimer translation
parallel to the bond orientation. The red-shifted defect frequencies,
that overlap into the bulk carbon range, are due to C-C partial rotation in
one instance, and in another due to dimer translation orthogonal to
the C-C bond.

$U_{3}$ in Fig. \ref{fig: qha-phonons}e is also notable for the
phase transition in strain evident \emph{via} the Gr{\"u}neisen discontinuity
at large dilation. The transition occurs between crystal volume 13.5
and 13.8 \r{A}\textsuperscript{3}/atom and lifts the trimer point group
symmetry from angular $C_{2v}$ to linear $C_{\infty v}$\@. This
change yields another unusually stiff mode at $43$ THz shown in Fig.
\ref{fig: qha-phonons}e, which is due to asymmetric stretching of
the C-C-C unit. Two other high-frequency localised modes exist, which have the same Gr{\"u}neisen parameters below the transition. At the transition, the lower frequency mode changes Gr{\"u}neisen sign, and begins increasing in frequency. This gives the appearance of a degeneracy at the 8th and 9th volumes in Fig. 8 e, but closer inspection reveals the modes are not degenerate. In Fig. 8 e we also note there are two low frequency localised modes in the gap evident below the transition. Above the transition there are still two localised modes at the gap, but they are not clearly observable in the total DOS plot due to proximity to bulk modes.

For $U_{4}$ in Fig. \ref{fig: qha-phonons}f, the stiffest defect
mode is due to the carbon interstitial stretching against its three
carbon nearest-neighbours. The observable red-shifted defect
mode, between the bulk carbon and zirconium bands, is due to the rigid
translation of the C\textsubscript{4 } unit which is made up of the
interstitial and it's three C nearest-neighbours.

Harmonic interatomic forces and vibration frequencies are related as
\begin{equation}
\left|\mathbf{\Phi}\right|=\left|\mathbf{\text{M}}\right|\prod_{i}\omega_{i}^{2}\,.\label{eq: determinent}
\end{equation}
In this expression the force constant matrix determinant  $|\mathbf{\Phi}|$ is equated with the product of the diagonal mass matrix entries $|\text{M}|$ and the squared frequencies $\omega_{i}^{2}$.

%\com{Define M as well as $\Phi$}
%\reply{19) M is the diagonal mass matrix. Definition added in text.}

The defective to perfect ratio $\left|\mathbf{\Phi}'\right|$/$\left|\mathbf{\Phi}\right|$
is a useful indicator of bond stiffness. In particular for stoichiometric
defects when $\left|\mathbf{\text{M}}\right|=\left|\mathbf{\text{M}}'\right|$
the force determinant ratio is a useful single-valued measure of the defect-induced
frequency redistribution. The ratio can be expressed as a density of states difference integral
\begin{equation}
\left|\mathbf{\Phi'}\right|/\left|\mathbf{\Phi}\right|=\text{exp}\,\int d\omega\,\left(g'(\omega)-g(\omega)\right)\,\text{ln}\,\omega\,,\label{eq: forceConstantRatio}
\end{equation}
which is how we have calculated the ratio in practice in this work.
%\com{Is that how you actually calculated it, rather than from the determinant of the force constant matrix? It is not clear in the text.}
%\reply{20) It is how it was  calculated, clarified now.}
It is also directly related to the classical excess vibrational entropy of the Frenkel defect by
\begin{align*}
S'-S = -\text{ln}\,\sqrt{\left|\mathbf{\Phi'}\right|/\left|\mathbf{\Phi}\right|}\,.
\end{align*}

%\com{I am suggesting to lighten the rather laborious discussion by introducing the classical excess entropy at this point, as I've inserted above. I commented out some original text below.}
%\reply{21) agree, good idea to lighten up.}

If $\left|\mathbf{\mathbf{\Phi'}}\right|/\left|\mathbf{\Phi}\right|<1$ we
infer net defect-induced crystal softening, else vice versa. For
Frenkel defects in ZrC we observe $\left|\mathbf{\Phi'}\right|/\left|\mathbf{\Phi}\right|\ll1$,
indicating stoichiometric defects reduce crystal stiffness.
This occurs despite the high-frequency vibrations of Frenkel
defects (visible in Fig. \ref{fig: qha-phonons}), that arise from locally stiffened interstitial carbon-carbon
bonds, as explained in what follows.

The values of $S'-S$ reported in Table \ref{tab: defect-Entropy-difference}
are positive which agrees  in sign with Frenkel pair entropies reported
for other materials.\citep{Gillan1983,Walsh2011,Jacobs2000,Sahni1982}
The unbound defects have larger vibrational entropies than bound defects,
which is also consistent with reports for other materials, \emph{e.g}.
the Frenkel pairs in In\textsubscript{2}O\textsubscript{3} reported
by Walsh \emph{et al}..\citep{Walsh2011}

$S'-S$ at fixed perfect-crystal volume ranges from $0.7$ $k_{\text{B}}$
to $2.3$ $k_{\text{B}}$d depending on the defect. Partial $S'-S$
contributions from carbon and zirconium in Table \ref{tab: defect-Entropy-difference}
show $S'-S$ is positive because the Zr-character modes soften more
than the carbon-type vibrations harden due to the localised interstitial
frequencies. An early Green's function analysis of Frenkel pairs in
CaF\textsubscript{2} by Gillan and Jacobs made a similar partition
of the defect entropy.\citep{Gillan1983} The conclusions they drew
were analogous \textendash{} Frenkel defect entropy $S'-S$ is net
positive, despite the locally negative entropy of the stiffened interstitial
vibrations.

\subsubsection*{Defect electronic characterisation}

Our DFT calculations predict that perfect ZrC has a density of states at the Fermi energy of  $\text{DOS}(E_{\text{F}})=0.13$ electron states/eV per atom. This value is similar to early estimates
by Ihara \emph{et al.\citep{Ihara1976}} with 0.09 states/eV per atom, and
Borukhovich and Geld\citep{Borukhovich1969} with 0.1 states/eV per
atom, and more recent LDA (GGA) calculations by Arya and Carter\citep{Arya2004} with 0.112 (0.129) states/eV. Absolute computed values depend somewhat on technical parameters. For example we observe a 5\% variation in $\text{DOS}(E_{\text{F}})$ on converging unitcell k-points between 4x4x4 and 40x40x40, +7\% change increasing $T_{\text{el}}=0.1$ eV to $T_{\text{el}}=0.2$ eV, +2\% change from Fermi-Dirac smearing to Methfessel-Paxton, and +12\% increase from GGA to LDA at fixed volume. Given these sensitivities, we next consider trends with defect type and thermal expansion rather than absolute values.

%\com{Would it be correct to add that these values are sensitive to the smearing scheme used in k-point sampling, so more important than absolute values are the trends when defects are introduced?}
%\reply{23)  Yes, I think so. Some notes:
%Ihara is exp. 
%Borukhovich is also exp.
%Ayra/Carter - 0.1 eV MP smearing, effective 7x7x7 unitcell kpoint sampling, giving 0.112/eV for LDA and 0.129/eV GGA 
%Tests to check DOS(EF) dependence on kpoints, LDA/GGA, FD/MP smearing, and smearing width, show that it is reasonable to suggest we should only look at trends. LDA-GGA %difference is fbiggest apparent effect. Text amended.
%}

The value of DOS(\emph{E}\textsubscript{F})
is modified by the presence and type of the Frenkel defect in the
crystal. A feature common to each defect is the tendency to increase DOS(\emph{E}\textsubscript{F})
compared to perfect ZrC. Compared to 0.13 states/eV/atom in perfect
ZrC, Frenkel defects increase DOS(\emph{E}\textsubscript{F}) by
between +0.02 states/eV/atom for $U_{3}$  and  +0.12 states/eV/atom
for $U_{2}$.

The value of DOS(\emph{E}\textsubscript{F}) in a transition metal
can become larger or  smaller with increasing Wigner-Seitz radius,
for example, this has been reported by Pettifor for the 4\emph{d}
series.\citep{Pettifor1977} In ZrC we find the precise
dependence of DOS(\emph{E}\textsubscript{F}) on volume expansion
varies depending on the specific defect configuration, illustrated
in Fig. \ref{fig: qha-el}. 

As lattice dilation and the insertion of defects both tend to increase
the value of DOS(\emph{E}\textsubscript{F}), electron entropy will tend to
stabilise defects. The distribution of DOS(\emph{E})
for each crystal configuration has a distinctive energy dependence
(see Fig. \ref{fig: qha-el}). At high temperatures a first-order Sommerfeld 
expansion will  be inadequate beyond qualitative inferences. 
Quantitative predictions including electronic defect free energies from the Mermin functional are
given subsequently.

\subsection{Defect thermodynamics\label{subsec:Defect-thermodynamics} }

\subsubsection{Enthalpy and entropy}

\begin{table}
\caption{Defect formation energy $\Delta U$ ($\Delta U=\Delta U_{0}+\Delta U_{\text{ZP}}+U_{\text{FS}}$)
and zero-point energy difference $\Delta U_{\text{ZP}}$. \label{tab:Intrinsic-defect-formation-energy}}
\begin{centering}
\begin{tabular}{ccc}
\hline 
Configuration & $\Delta U$ (eV/defect) & $\Delta U_{\text{ZP}}$ (eV/defect)\tabularnewline
\hline 
\hline 
$B_{2}$  & 3.734  & -0.031 \tabularnewline
$B_{3}^{\text{}}$  & 3.247  & -0.016 \tabularnewline
$U_{2}$  & 4.436  & -0.042\tabularnewline
$U_{3}^{\text{}}$  & 4.335  & -0.027 \tabularnewline
$U_{4}$  & 4.441  & -0.030 \tabularnewline
\hline 
\end{tabular}
\par\end{centering}
\end{table}

\begin{figure}
\begin{raggedright}
\includegraphics[scale=0.55]{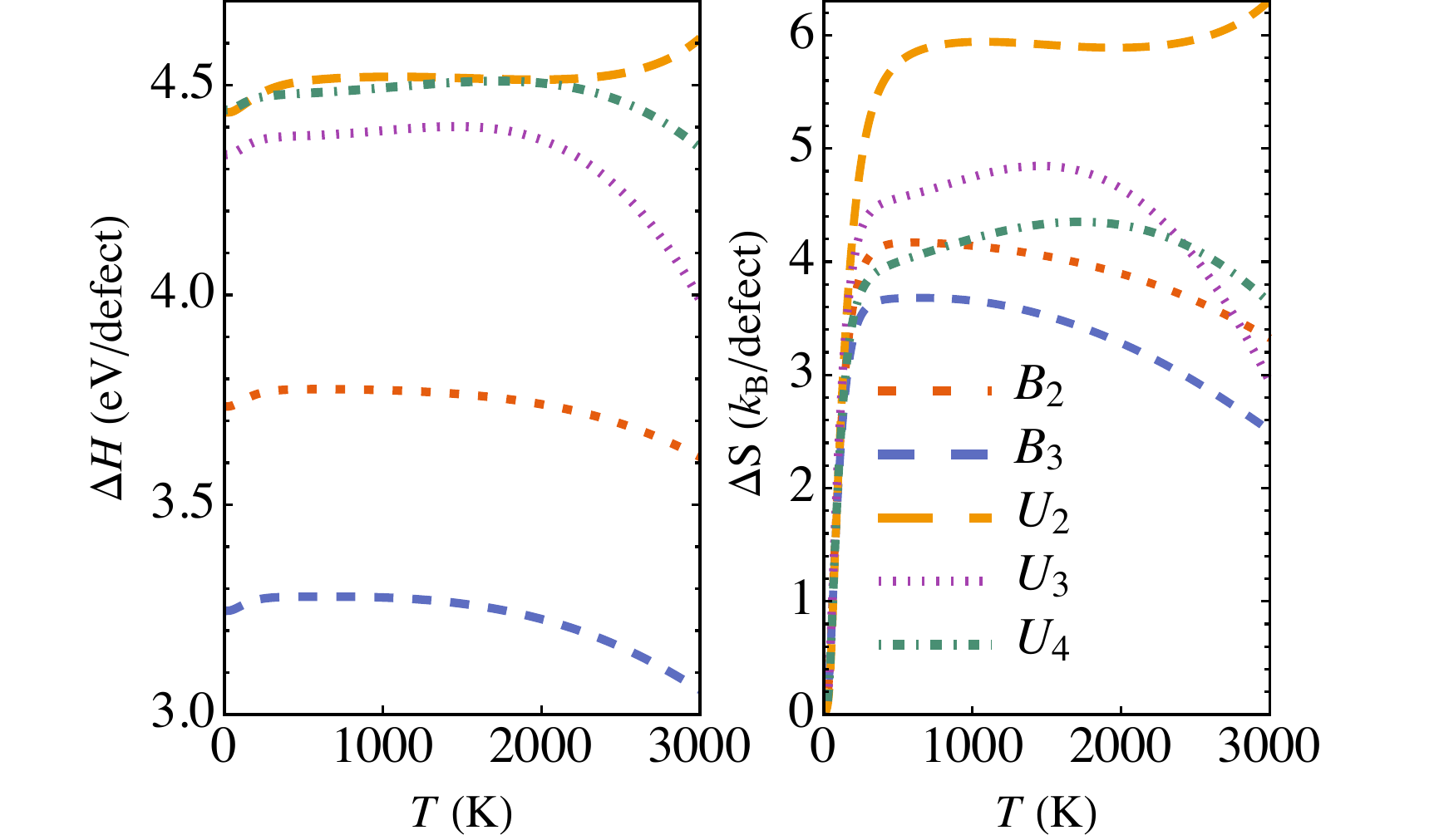}
\par\end{raggedright}
\caption{Ambient pressure formation energy $\Delta H=H^{\text{defect}}-H^{\text{perfect}}$
and entropy $\Delta S=S^{\text{defect}}-S^{\text{perfect}}$. Frenkel
ZrC has one defect per supercell. \label{fig: relativeThermodynamics-1} }
\end{figure}

The $B_{3}$ Frenkel defect is the most energetically stable configuration at
$3.2$ eV/defect, followed by $B_{2}$ at $3.7$ eV/defect. Unbound
pairs are less stable, costing between $4.3$ and $4.5$ eV/defect
to introduce to the lattice. The full set of values at $T=0$ K is
listed in Table \ref{tab:Intrinsic-defect-formation-energy}, and
shown as a function of temperature at ambient pressure in Fig. \ref{fig: relativeThermodynamics-1}.
The defect formation enthalpies, $\Delta H=H^{\text{defect}}-H^{\text{perfect}}$,
show  variations of a few tenths of an eV over $3000$ K. The
relatively weak temperature dependence is due to partial cancellation of vibrational and electronic contributions.  To see this consider the bound dimer enthalpy $\Delta H(B_3)$ as a typical example. Between 0 K and 3000 K, the electronic part increases by $\Delta H_\text{el}(B_3)=0.14 $ eV/defect, and quasiharmonic part decreases by $\Delta H_\text{qh}(B_3)=-0.25 $ eV/defect. Overall change is modest at $\Delta H(B_3)=-0.12 $ eV/defect, as evident in Fig. \ref{fig: relativeThermodynamics-1}.

%\com{Point to where you have shown this effect}.
%\reply{24) Numerical example to show cancellation effect given in text above.}

$B_{3}$ has the smallest ambient pressure formation entropy and $U_{2}$
has the largest, as seen in Fig. \ref{fig: relativeThermodynamics-1}.
The ordering of defect entropies agrees with the rudimentary
estimates in Table \ref{tab: defect-Entropy-difference}
computed from fixed-perfect-volume force-constant determinants, though the ambient pressure values are larger.

\subsubsection{Gibbs energy of defect formation}

\begin{figure}
\begin{centering}
\includegraphics[scale=0.72]{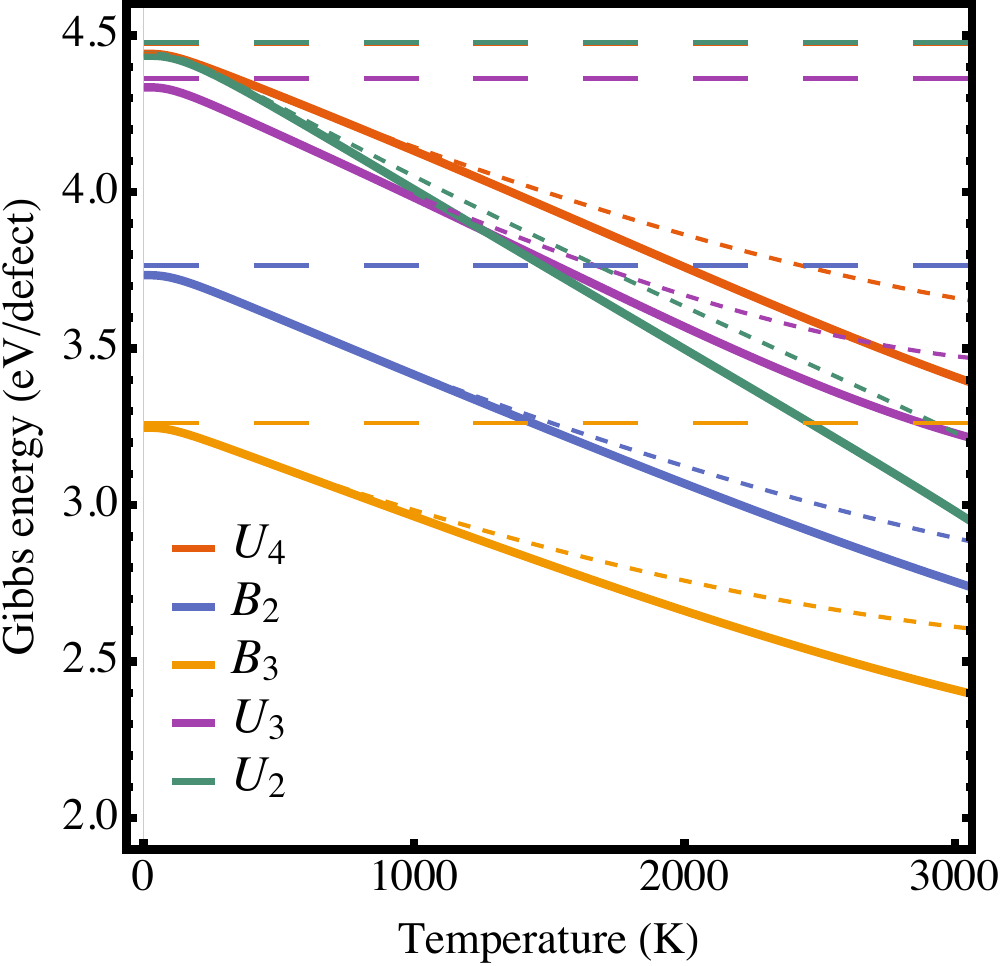}
\par\end{centering}
\caption{
Defect Gibbs formation energy (Eqn. \ref{eq: gibbsEnergyFS}) for each defect type. \textit{Solid} includes quasiharmonic and electronic contributions, \textit{short-dashed} includes quasiharmonic, and \textit{long-dashed} is the zero-temperature formation energy without zero-point contributions.}
%i) \emph{Solid line} is $E_{0}+E_{\text{FS}}+F_{\text{qh}}+F_{\text{el}}$
%level-of-theory Gibbs free energies. ii) \emph{Short-dash} is $E_{0}+E_{\text{FS}}+F_{\text{qh}}$
%evel. iii) \emph{Long-dash} is $E_{0}+E_{\text{FS}}$ level (energy
%including finite-size corrections but not zero point energy).
\label{fig: Gibbs-free-energy-of-defect-formation}
%\reply{26) Equations removed from caption, and referenced to definition in method section.}

\end{figure}

The Gibbs energy of defect formation is shown in Fig. \ref{fig: Gibbs-free-energy-of-defect-formation}
up to 3000 K (above this temperature anharmonic thermodynamic contributions
become considerable in ZrC)\citep{Duff2015}. Thermally excited electrons
and phonons can reduce the energy to form a Frenkel defect by more than
$1$ eV over the interval $0$ K to $3000$ K. This is mostly due to phonons but the
electron contribution gains relative importance with temperature. In terms of the Gibbs energy of defect formation, zero-point effects are negligible. This is observable in Fig. \ref{fig: Gibbs-free-energy-of-defect-formation} and Table. \ref{tab:Intrinsic-defect-formation-energy}.

\subsubsection{Bulk modulus}
\begin{center}
\begin{figure*}
\begin{raggedright}
\includegraphics[scale=0.42]{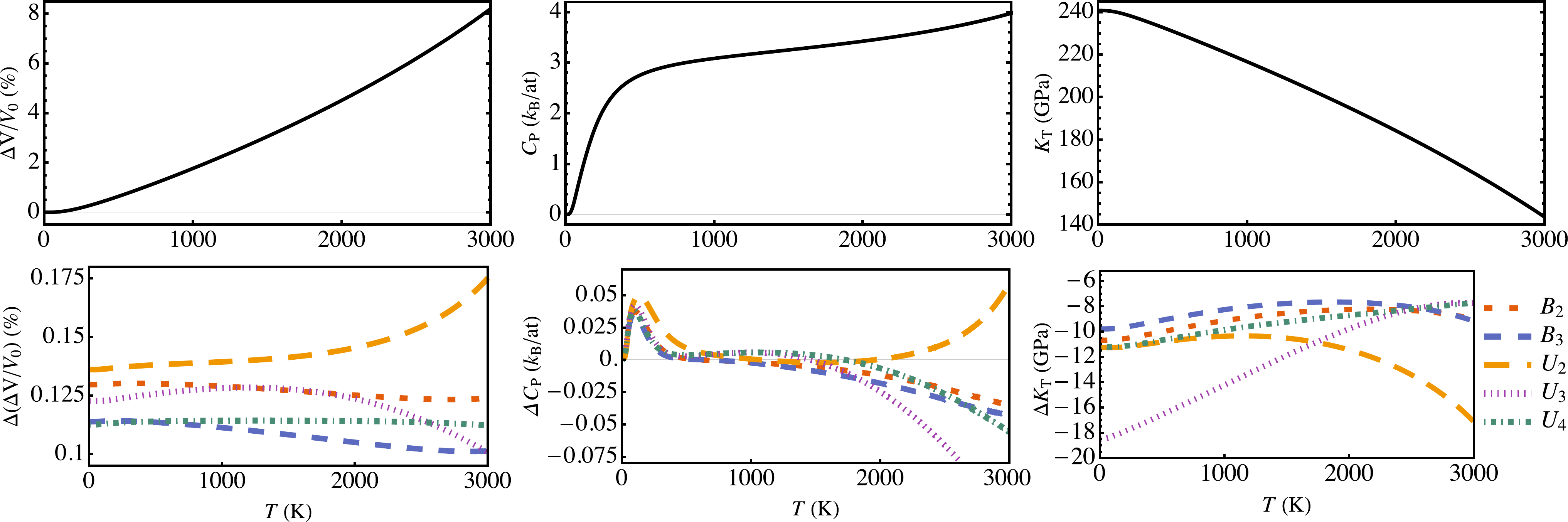}
\par\end{raggedright}
\caption{\emph{Top}, \emph{left to right}: Thermal volume expansion ($\Delta V/V_{0}$),
isobaric heat capacity ($C_{\text{P}}$), and isothermal bulk modulus
($K_{\text{T}}$) for perfect ZrC. \emph{Bottom}: The defective minus
perfect difference, $\Delta\mathcal{O}=\mathcal{O}^{\text{defect}}-\mathcal{O}^{\text{perfect}}$,
for each quantity. The defective  ZrC here has one Frenkel defect per supercell.
\label{fig: qha-el-1}}
\end{figure*}
\par\end{center}

\begin{table}
\caption{Isothermal bulk moduli in units of GPa for perfect and defective ZrC.
Frenkel ZrC has one defect per supercell. Selected calculated\citealp{Jiang2017,Ivashchenko2009,Cheng2004a,Jochym2000,Xie2016}
and measured\citep{Chang1966} values are included. \label{tab: isothermal-bulkmod}}

\begin{tabular}{ccccc}
\hline 
Configuration & $K_{0}$  & $K_{300}$  & $K_{1000}$ & $K_{3000}$\tabularnewline
\hline 
\hline 
perfect & 240.7 & 236.2 & 216.7  & 143.6 \tabularnewline
$B_{2}$  & 230.0  & 225.8  & 207.6  & 134.6 \tabularnewline
$B_{3}^{\text{}}$  & 230.9  & 226.8  & 208.4  & 134.4 \tabularnewline
$U_{2}$  & 229.5 & 225.1 & 206.3 & 126.4 \tabularnewline
$U_{3}^{\text{}}$  & 222.2 & 218.8 & 202.5 & 135.9 \tabularnewline
$U_{4}$  & 229.5  & 225.2 & 206.8  & 135.9 \tabularnewline
\multirow{2}{*}{ZrC (PBE)} & 232.2\citep{Jiang2017}, 219\citep{Ivashchenko2009}, & \multirow{2}{*}{-} & \multirow{2}{*}{-} & \multirow{2}{*}{-}\tabularnewline
 & 232\citep{Cheng2004a,Jochym2000}, 234\citep{Xie2016} &  &  & \tabularnewline
ZrC\textsubscript{0.94}(ultra-sonic) & - & 223\citep{Chang1966} & - & -\tabularnewline
\hline 
\end{tabular}
\end{table}

We estimate the isothermal bulk modulus as

\begin{equation}
K_{\text{T}}=-V\frac{\partial^{2}G}{\partial V^{2}}\,.\label{eq: isothermalBulkModulus}
\end{equation}
$K_{\text{T}}$ is shown in Fig. \ref{fig: qha-el-1}, and specific
values are reported in Table \ref{tab: isothermal-bulkmod}. 

$K_{\text{T}}$ is known experimentally for ZrC from pulsed-ultrasonic
measurements by Chang and Graham, who report  $K_{\text{300}}^{\text{exp}}=223$
GPa for ZrC\textsubscript{0.94}. At 300 K, we predict $K_{\text{300}}^{\text{perfect}}=236$
GPa, and $K_{\text{300}}^{\text{defect}}\approx219-227$ GPa depending
on the type of Frenkel defect present in the simulation cell. The
defect reductions in $K_{\text{T}}$ are between $5-10$ \% at the Frenkel concentration of  one defect by supercell (1/32 C at. \%). Frenkel-induced $K_{\text{T}}$ variation  is relevant to ZrC applications in high radiation and temperature environments. ZrC is radiation tolerant and has a relatively low neutron scattering cross-section $\sigma_\text{ZrC} \approx 0.2\,\sigma_\text{ZrAlloy}$),\citep{Azevedo2011, Jiang2013, Jiang2017} but there is reportedly less known about ZrC for fuel clad design that comparable materials such as SiC.\citep{Katoh2012, Azevedo2011, Snead2007} First principles data on Frenkel-induced changes to structural parameters is therefore likely useful for sensitivity analyses in multiphysics simulations of accident tolerant fuel materials.\citep{Marino2015, Geelhood2014, Liu2018, Devanathan2010}

%\com{Can you support this statement with a reference from the engineering literature?}
%\reply{27) The original statement was that a 10\% in bulk mod was relevant to engineering. I found that difficult to precisely support  with a citation. I've added a couple of sentences that are citeable though. }

The $K_{0}$ value we predict for perfect ZrC differs from other computational
studies by 10-20 GPa.\citep{Jiang2017,Ivashchenko2009,Cheng2004a,Xie2016}
While most studies use the PBE GGA functional, we use the LDA exchange-correlation.
Our choice of the LDA functional has been motivated by the adequate performance
of LDA and poor performance of PBE at high temperature for ZrC reported
by Duff \emph{et al.} \citep{Duff2015}.

\subsubsection{Heat capacity}

The ambient-pressure isobaric heat capacity,

\begin{align}
C_{\text{P}} & =-T\frac{\partial^{2}G}{\partial T^{2}}\,\label{eq: heatCapacity}\\
 & =C_{\text{V}}+T\frac{\partial S}{\partial V}\frac{\partial V}{\partial T}\,,\label{eq: heatCapacity2}
\end{align}
is shown for perfect ZrC in Fig. \ref{fig: qha-el-1}, along with
the excess defect contribution
$\Delta C_{\text{P}}(T)=C_{\text{P}}^{\text{defect}}(T)-C_{\text{P}}^{\text{perfect}}(T)$ associated with forming a single defect.
%\com{define excess quantity}\reply{28) done}

%Note the striking peak in the excess heat capacity of the defective crystal at low temperature.
%This is a rather nice illustration of quantum mechanics at several
%hundred Kelvin in the elementary excitations of a garden-or-common
%structural ceramic. 
The low-temperature peak arises from the differential temperature
occupation of the perfect and defect crystal phonon spectra. Low frequency
modes are softer in the defective lattice compared to the perfect
crystal. As temperature increases, the defect crystal heat capacity
increases more than perfect ZrC resulting in the peak. As the crystal approaches
the Debye temperature, quantum effects in the heat capacity fall off
and the difference $\Delta C_{\text{P }}(T)$ recedes. 
We emphasise the low-temperature bump is not an isobaric effect, \emph{i.e}., it is not due to the second term in
Eqn. \ref{eq: heatCapacity2}. It is also present 
in a plot of the constant volume equivalent $\Delta C_{\text{V}}(T)$.

\begin{table}
\caption{Volume for perfect and defective ZrC at 0 K, 300 K, 1000 K and 3000
K. Frenkel ZrC has one defect per supercell. Units of \r{A}\protect\textsuperscript{3}/atom
\label{tab: alatt}}

\begin{tabular}{ccccc}
\hline 
Configuration & $V_{0}$  & $V_{300}$  & $V_{1000}$  & $V_{3000}$ \tabularnewline
\hline 
\hline 
perfect & 12.706 & 12.740 & 12.931 & 13.749\tabularnewline
$B_{2}$  & 12.836 & 12.870 & 13.060 & 13.873\tabularnewline
$B_{3}^{\text{}}$  & 12.820 & 12.854 & 13.042 & 13.850\tabularnewline
$U_{2}$  & 12.843 & 12.877 & 13.070 & 13.924\tabularnewline
$U_{3}^{\text{}}$  & 12.829 & 12.864 & 13.059 & 13.850\tabularnewline
$U_{4}$  & 12.819 & 12.854 & 13.045 & 13.861\tabularnewline
ZrC (PBE) & 12.903\citep{Jochym2000} 12.887\citep{Xie2016}  & - & - & -\tabularnewline
ZrC\textsubscript{0.94}(exp.) & - & 12.973\citep{Savvatimskiy2017}  & - & -\tabularnewline
ZrC\textsubscript{0.95}(exp.) & - & 12.928\citep{Chang1966}   & - & -\tabularnewline
\hline 
\end{tabular}

\end{table}

\subsubsection{Thermal expansion}

The isotropic volume expansion $\frac{\Delta V}{V}$ is found from the volume that
minimizes the Gibbs free energy at each temperature,
%\begin{equation}
%\frac{\partial G}{\partial V}=0\,.\label{eq: thermalExpansion}
%\end{equation}
It is shown for each configuration
in Fig. \ref{fig: qha-el-1}, at a concentration of one defect per
64 atom supercell ($n_{\text{fp}}=3$ C at. \%), and values are listed
in Table \ref{tab: alatt}. The volume expansion induced by one Frenkel defect in the supercell is approximately 0.12\%, with weak dependence on temperature. We consider in the following sections the effect of a thermal population of defects.
\begin{figure}
\begin{centering}
\includegraphics[scale=0.62]{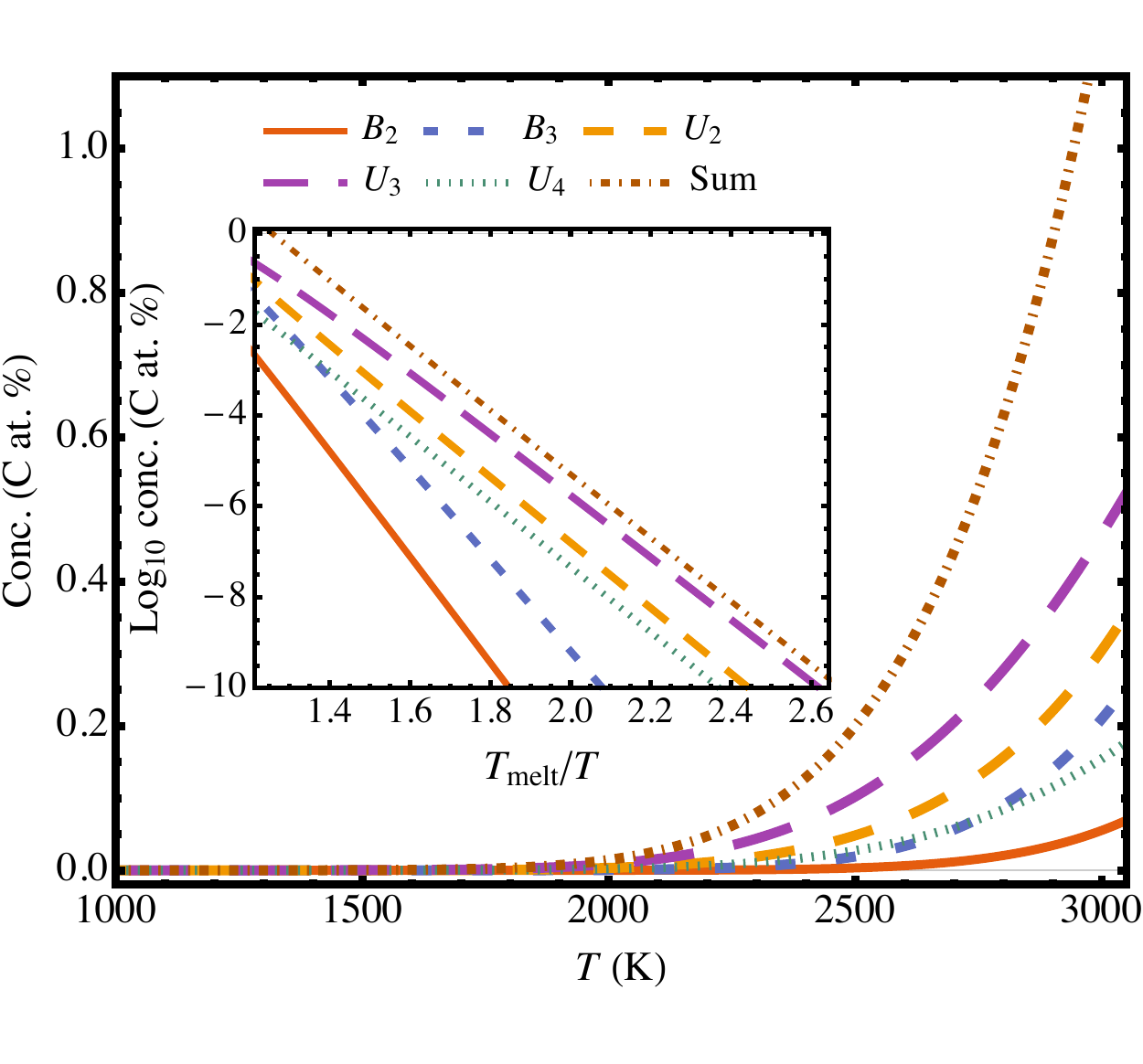}
\par\end{centering}
\caption{Frenkel pair concentration. Log scale inset.\label{fig:Defect-concentration-in-ZrC}}
\end{figure}

\begin{figure}
\begin{centering}
\includegraphics[scale=0.80]{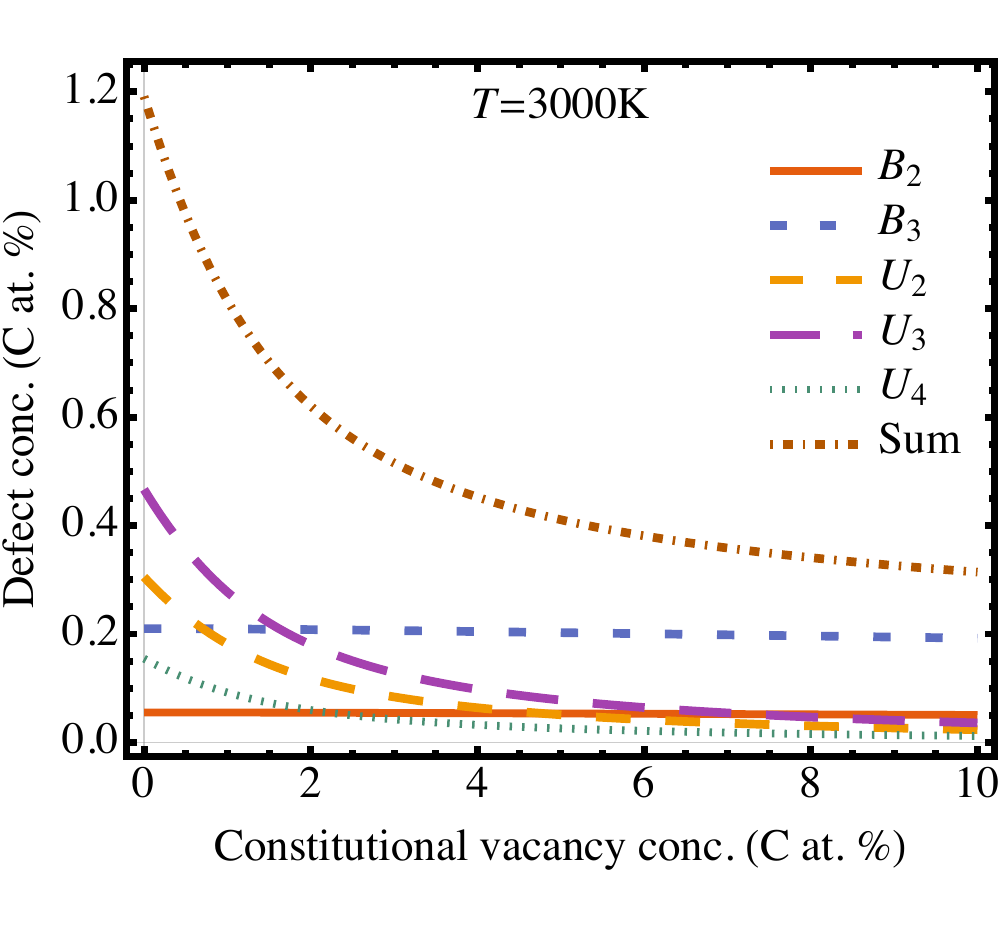}
\par\end{centering}
\caption{Coupling of equilibrium Frenkel defect concentration at 3000 K to carbon substoichiometry.
Constitutional carbon vacancies weakly suppress bound pair defects and
strongly suppress unbound pairs. \label{fig: frenkelConc-vs-vacConc}}
\end{figure}

\begin{table}
\caption{Frenkel pair concentration.  Concentrations measured in C at. \%.\label{tab: temperature-vacancyConc-frenkelConc}}
\centering{}%
\begin{tabular}{ccccc}
\hline 
\noalign{\vskip\doublerulesep}
$T$ (K)  & $n_{\text{vac}}^{\text{C}}$  & $n_{\text{fp}}^{\text{bound}}$  & $n_{\text{fp}}^{\text{unbound}}$  & $n_{\text{fp}}^{\text{total}}$ \tabularnewline[\doublerulesep]
\hline 
 & 0 & $5\times10^{-4}$ & 0.01 & 0.01\tabularnewline
 & 2 & $5\times10^{-4}$ & $9\times10^{-5}$ & $6\times10^{-4}$\tabularnewline
2000 & 5 & $5\times10^{-4}$ & $4\times10^{-5}$ & $5\times10^{-4}$\tabularnewline
 & 10 & $5\times10^{-4}$ & $2\times10^{-5}$ & $5\times10^{-4}$\tabularnewline
\cline{2-5} 
 & 0 & 0.02 & 0.18 & 0.20\tabularnewline
 & 2 & 0.02 & 0.02 & 0.04\tabularnewline
2500 & 5 & 0.02 & $6\times10^{-3}$ & 0.03\tabularnewline
 & 10 & 0.02 & $3\times10^{-3}$ & 0.02\tabularnewline
\cline{2-5} 
 & 0 & 0.27 & 0.93 & 1.19\tabularnewline
 & 2 & 0.26 & 0.36 & 0.62\tabularnewline
3000 & 5 & 0.26 & 0.16 & 0.41\tabularnewline
\multirow{1}{*}{} & 10 & \multirow{1}{*}{0.24} & \multirow{1}{*}{0.07} & \multirow{1}{*}{0.31}\tabularnewline
\hline 
\end{tabular}
\end{table}

\subsubsection{Defect concentration}

$n_{\text{fp}}$ is shown in Fig. \ref{fig:Defect-concentration-in-ZrC}
between $1000$ and $3000\,$K,
%\com{ Fig. 9  only shows $3000\,$K so I've reworded this.}
%\reply{29) I modified this figure recently but not the associated text, thanks for catching.}
calculated with the ideal solution model. 
Details are provided in Appendix \ref{sec:idealSolnModel}. 
%In summary
%the value of $n_{\text{fp}}$ we report in ZrC is high. 
$n_{\text{fp}}$ ranges from 0.01\% at $2000$ K (0.54 $T_{\text{m}}$) to $1.2$
C at.\% at $3000$ K (0.81 $T_{\text{m}}$), where the experimental melting point is  $T_{\text{m}}=3700\,$K. We are unaware
of any reports of such high concentrations of thermally induced self-interstitials in any material hundreds of degrees below the melting point. Recent evidence, both experimental and with molecular dynamics,  has indicated the existence of thermally induced self-interstitials in aluminium, but only within $2-4\,$K of $T_{\text{m}}$\citep{Safonova2016}, and at a concentration two or three times below that of the thermally induced vacancies. Since we are dealing here with a compound, the concept of thermally induced vacancies is not directly relevant; in the case of ZrC any concentration of carbon vacancies not associated with Frenkel defect formation is associated with non-stoichiometry. In practice, the ZrC lattice structure supports a high level of carbon deficit, manifest as vacancies on the carbon sublattice. The law of mass action dictates that such constitutional vacancies  will reduce the concentration of unbound interstitials. Fig. \ref{fig: frenkelConc-vs-vacConc} shows the calculated effect of  substoichiometry on the Frenkel defect population at $3000\,$K. Some specific values are also listed in Table \ref{tab: temperature-vacancyConc-frenkelConc}.

The exceptional population of Frenkel defects is explained by the following five points:
\begin{enumerate}
\item ZrC supports at least five symmetry-inequivalent Frenkel-pair types,
\{$B_{2}$, $B_{3}$, $U_{2}$, $U_{3}^{\text{}}$, $U_{4}$\}.
\item Interstitial carbon tends to occupy low-symmetry sites. The associated
multiplicities are listed in Table \ref{tab: Identified-stable-configurations}. 
\item Frenkel pairs increase the electronic density of states at the Fermi
level. 
\item Frenkel pairs expand the lattice, with a net softening of vibration frequencies.
\item ZrC remains crystalline to an unusually high melting temperature.
Configurational, electronic, and phonon entropies, which are all stabilising
for defects, therefore have uncommonly large stabilising effects.
\end{enumerate}
%Up to this point predictions have focused on Frenkel pairs in stoichiometric
%ZrC. ZrC is reliably reported to support a wide range of sub-stoichiometry on
%the carbon sub-lattice. Sub-stoichiometry reduces the number of sites
%at which Frenkel pairs may form, and increases the number of lattice
%sites at which an interstitial may annihilate with a vacancy. The
%effect of constitutionally vacant carbon sites on $n_{\text{fp}}$
%is shown in Table \ref{tab: temperature-vacancyConc-frenkelConc},
%and Fig. \ref{fig: frenkelConc-vs-vacConc}. 

%The carbon composition dependence is weak for bound pairs, and stronger
%for unbound pairs, but each in case constitutional vacancies suppress
%Frenkel concentration. Consider an explicit numerical example. In
%stoichiometric ZrC, $n_{\text{fp}}=1.2$ C at.\% at $3000$ K. With
%a vacancy concentration of $10$ C at.\%, overall Frenkel concentration
%decreases from $n_{\text{fp}}=1.2$ to $0.3$ C at.\%, \emph{via}
%a 8 \% reduction in bound pairs and 92 \% reduction in unbound pair
%concentrations. For additional values refer to Table \ref{tab: temperature-vacancyConc-frenkelConc}
%and Fig. \ref{fig: frenkelConc-vs-vacConc}. 

\section{Conclusions\label{sec:Conclusions}}

In simulations of a ZrC crystal, five distinct Frenkel defect configurations have been found that are metastable
at zero temperature. Two of the configurations are bound pairs and
in three other configurations the vacancy and interstitial are separated.
While at 300 K the equilibrium concentration of Frenkel pairs is negligible,  at $3200\,$K they were 
frequent enough to occur spontaneously during a first-principles molecular dynamics simulation.
%low at
%$n_{\text{fp}}\approx10^{-34}$ \%, approximately one Frenkel pair
%per few billion tonnes of ZrC. 

We have performed quasi-harmonic lattice dynamics calculations to study further the thermodynamics of these defects.
The results indicate surprisingly high concentrations of bound and unbound Frenkel pairs:  in the stoichiometric crystal we estimate
0.01 \%  per mole at 2000 K, 0.2 \% at 2500 K, and 1.2 \% at 3000 K. For a substoichiometric crystal we can predict how much the number would
be suppressed by recombination. The high proportion of bound vacancy-interstitial pairs maintains the high concentration of Frenkel pairs even in the presence of constitutional vacancies. We find for ZrC$_{0.9}$ at 3000 K the concentration is reduced from the stoichiometric value  1.2 \% to 0.3 \%. This is by reducing the concentration of bound pairs from 0.93 \% to 0.07 \%, and the unbound pairs from 0.27 \% to 0.24 \%.

The high concentration of Frenkel defects change the properties of ZrC several important ways. A concentration of one defect per 64 atom ZrC cell approximately doubles the electronic density of states at the Fermi level, decreases the bulk modulus by 8-18 GPa, and dilates the lattice by 0.1-0.15 \%. The Frenkel contribution to the heat capacity is relatively small, not exceeding 0.05 $k_\text{B}$ for $T<2000$ K.

%\com{More detailed conclusions needed}.
%\reply{29) Added}

\cleardoublepage{}

\section{References}

\bibliographystyle{unsrt}
\bibliography{Intrinsic-defects-ZrC-manuscript}

\cleardoublepage{}

\section{Acknowledgements}

T.A.M. and M.W.F. gratefully acknowledge the financial support of
EPSRC Programme Grant (Grant No. EP/K008749/1) Material Systems for
Extreme Environments (XMat), and is grateful to the UK Materials and
Molecular Modelling Hub for computational resources, which is partially
funded by EPSRC (EP/P020194/1). This work was funded under the embedded
CSE 33 of the ARCHER UK National Supercomputing Service (\href{http://www.archer.ac.uk}{http://www.archer.ac.uk}).
A.I.D. would like to thank the STFC Hartree Centre for supporting
this effort by allowing human resource to be dedicated to this work.
This work was in part supported by the STFC Hartree Centre\textquoteright s
Innovation: Return on Research programme, funded by the UK Department
for Business, Energy \& Industrial Strategy.

\cleardoublepage{}

\appendix

\section*{Appendix}

%\com{Do you intend to submit this as supplementary material, a separate document from the paper?} 
%\reply{I prefer as an appendix  rather than being a separate document. Separate document is okay for data, but it is a hassle sometimes because you have to click on a separate link, rather than just scroll to the end of the document, but if you or the journal have some preference, that is fine.}

\section{Ideal solution model\label{sec:idealSolnModel}}

%\com{Do you mean to ignore all but one interstitial type in this, it would be straightforward allthough more complicated to include all the interstitials $B2,B3, U2...$ etc. but perhaps not necessary.}
%\reply{For the sake of derivation, I think it is sufficient only to distinguish the main types. The equations for sub-types can be easily inferred. For completeness I'll state the five sub-type equations at the end, but only include the bound and unbound in derivation.}

%

We derive an ideal solution model for ZrC based on the configurational
entropy of the carbon sub-lattice. The articles of interest on the
lattice are constitutional carbon vacancies (denoted $N_{1}$), bound
Frenkel pairs ($N_{2}$), and unbound Frenkel pairs ($N_{3}$). The
basis for counting combinations of these species is the number of
perfect sites $\left(N-N_{1}-N_{2}-2N_{3}\right)$, the number of
vacancies $\left(N_{1}+N_{3}\right)$, bound pairs ($N_{2}$) and
free interstitials ($N_{3}$). The number of combinations or unordered
selections without repetition is

\[
Z^{\text{}}=m_{2}^{N_{2}}m_{3}^{N_{3}}\frac{N!}{\left(N-N_{1}-N_{2}-2N_{3}\right)!\,\left(N_{1}+N_{3}\right)!\,N_{2}!\,N_{3}!}\,.
\]
Factors $m_{2}$ and $m_{3}$ account for the degeneracy of interstitial
sites for bound pairs and unbound pairs, with $m_{2}$ counted with respect to carbon vacancy sites and $m_{3}$ with respect to carbon perfect lattice sites.

The first-order Stirling approximation is applied to the lattice configurational
entropy 

\[
S^{\text{mix}}=\text{ln}\,Z\,,
\]
giving the expression

\begin{align*}
S^{\text{mix}} & =N_{2}\text{ln}\,m_{2}+N_{3}\text{ln}\,m_{3}+N\text{ln}\,N\\
 & \,\,\,-(N-N_{1}-N_{2}-2N_{3})\text{ln}\,(N-N_{1}-N_{2}-2N_{3})\\
 & \,\,\,-\left(N_{1}+N_{3}\right)\text{ln}\,\left(N_{1}+N_{3}\right)-N_{2}\text{ln}\,N_{2}-N_{3}\text{ln}\,N_{3}\,.
\end{align*}

The Gibb's free energy of the system of defects

\[
\sum_{i=1}^{i=3}N_{i}\Delta G_{i}=TS^{\text{mix}}\,,
\]
is minimised with respect to the number of each type of defect

\[
\Delta G_{i}=T\partial_{N_{i}}S^{\text{mix}}\,.
\]
For bound pairs the configuration entropy term is

\[
\partial_{N_{2}}S^{\text{mix}}=\text{ln}\,\frac{m_{2}\,\left(N-N_{1}-N_{2}-2N_{3}\right)}{N_{2}}\,,
\]
and for unbound pairs

\[
\partial_{N_{3}}S^{\text{mix}}=\text{ln}\frac{m_{3}\,\left(N-N_{1}-N_{2}-2N_{3}\right){}^{2}}{\left(N_{1}+N_{3}\right)N_{3}}\,.
\]

The free energies to form bound and unbound pairs are given by

\[
\Delta G_{2}=T\,\text{ln}\,\frac{m_{2}\left(N-N_{1}-N_{2}-2N_{3}\right)}{N_{2}}\,,
\]
and

\[
\Delta G_{3}=T\,\text{ln}\,\frac{m_{3}\left(N-N_{1}-N_{2}-2N_{3}\right){}^{2}}{\left(N_{1}+N_{3}\right)N_{3}}\,.
\]

The concentrations ($n_{i}=N_{i}/N$) of each defect type are

\[
n_{2}=(1-n_{1}-n_{2}-2n_{3})m_{2}\,\text{exp}\left(-\frac{\Delta G_{2}}{T}\right)\,,
\]
and

\[
n_{3}=\frac{\left(1-n_{1}-n_{2}-2n_{3}\right){}^{2}}{\left(n_{1}+n_{3}\right)}\,m_{3}\,\text{exp}\left(-\frac{\Delta G_{3}}{T}\right)\,.
\]

For completeness the concentration of each sub-type of defect is stated
explicitly. For bound pairs, the concentrations of dimer and trimer
configurations are
\begin{widetext}
\[
n_{B_{2}}=\left[1-n_{1}-n_{B_{2}}-n_{B_{3}}-2\left(n_{U_{2}}+n_{U_{3}}+n_{U_{4}}\right)\right]\,m_{B_{2}}\,\text{exp}\left(-\frac{\Delta G_{B_{2}}}{T}\right)\,,
\]

\[
n_{B_{3}}=\left[1-n_{1}-n_{B_{2}}-n_{B_{3}}-2\left(n_{U_{2}}+n_{U_{3}}+n_{U_{4}}\right)\right]\,m_{B_{3}}\,\text{exp}\left(-\frac{\Delta G_{B_{3}}}{T}\right)\,\,,
\]
and the concentrations of unbound Frenkel pairs of dimer, trimer and
tetramer configurations are

\[
n_{U_{2}}=\frac{\left[1-n_{1}-n_{B_{2}}-n_{B_{3}}-2\left(n_{U_{2}}+n_{U_{3}}+n_{U_{4}}\right)\right]{}^{2}}{\left(n_{1}+n_{U_{2}}+n_{U_{3}}+n_{U_{4}}\right)}\,m_{U_{2}}\,\text{exp}\left(-\frac{\Delta G_{U_{2}}}{T}\right)\,,
\]

\[
n_{U_{3}}=\frac{\left[1-n_{1}-n_{B_{2}}-n_{B_{3}}-2\left(n_{U_{2}}+n_{U_{3}}+n_{U_{4}}\right)\right]{}^{2}}{\left(n_{1}+n_{U_{2}}+n_{U_{3}}+n_{U_{4}}\right)}\,m_{U_{3}}\,\text{exp}\left(-\frac{\Delta G_{U_{3}}}{T}\right)\,,
\]

\[
n_{U_{4}}=\frac{\left[1-n_{1}-n_{B_{2}}-n_{B_{3}}-2\left(n_{U_{2}}+n_{U_{3}}+n_{U_{4}}\right)\right]{}^{2}}{\left(n_{1}+n_{U_{2}}+n_{U_{3}}+n_{U_{4}}\right)}\,m_{U_{4}}\,\text{exp}\left(-\frac{\Delta G_{U_{4}}}{T}\right)\,.
\]
\end{widetext}

\section{Electron states\label{sec:Electron-states}}

Fig. \ref{fig: qha-el} in the main text shows the electron density
of states for each defect on the interval {[}-0.5, 0.5{]} eV. The
density of states is provided for interval {[}-6, 6{]} eV in Fig.
\ref{fig: qha-el-large}. 
\begin{center}
\begin{figure*}
\begin{raggedright}
\includegraphics[scale=0.55]{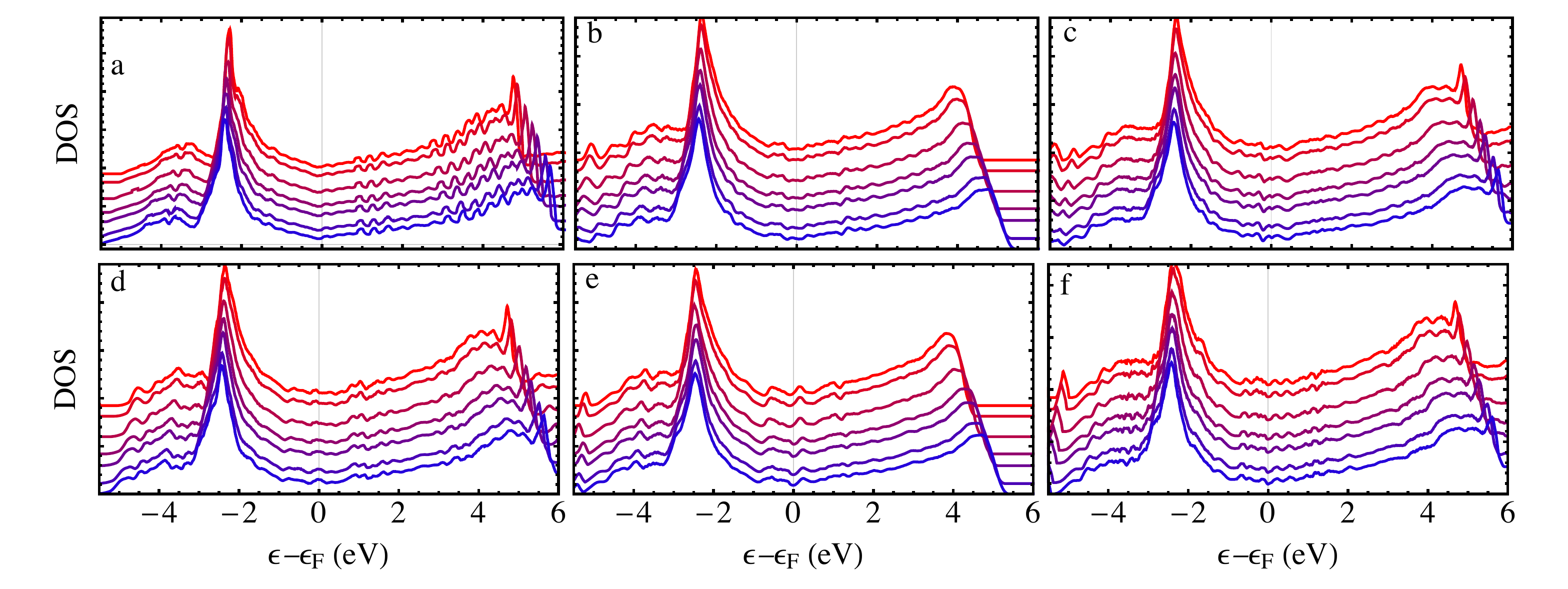}
\par\end{raggedright}
\caption{Electron density of states as a function of dilation. a) Perfect ZrC,
b) $B_{2}$ bound dimer, c) $B_{3}$ bound trimer, d) $U_{2}$ unbound
dimer, e) $U_{3}^{\text{}}$ unbound trimer, f) $U_{4}$ unbound tetramer.
Thermal expansion increases from blue to red. Consecutive densities
of states are shifted up proportional to the increase in volume. \label{fig: qha-el-large}}
\end{figure*}
\par\end{center}

\cleardoublepage{}
\end{document}